\documentclass{aa}  
\usepackage{natbib}
\usepackage{subcaption}
\usepackage{graphicx}
\usepackage{txfonts}
\usepackage{hyperref}
\providecommand{\bjdtdb}{\ensuremath{\rm {BJD_{TDB}}}}

\providecommand{\mj}{\ensuremath{\,M_{\rm J}}}
\providecommand{\rj}{\ensuremath{\,R_{\rm J}}}

\providecommand{\fave}{\langle F \rangle}
\providecommand{\fluxcgs}{10$^9$ erg s$^{-1}$ cm$^{-2}$}

\begin{document} 

    \title{TraMoS}

   \subtitle{V. Updated ephemeris and multi-epoch monitoring of the hot Jupiters WASP-18Ab, WASP-19b, and WASP-77Ab} 
   
   \author{P\'ia Cort\'es-Zuleta\inst{1,2,5}
            \and
          Patricio Rojo\inst{1}
          \and
          Songhu Wang\inst{2,6}
          \and
          Tobias C. Hinse\inst{3}
          \and
          Sergio Hoyer\inst{4}
          \and
          Bastian Sanhueza\inst{1}
          \and
          Patricio Correa-Amaro\inst{1}
          \and
          Julio Albornoz\inst{1}\fnmsep
          }

   \institute{Departamento de Astronom\'ia, Universidad de Chile, 
            Camino El Observatorio 1515, Las Condes, Santiago, Chile\\
              \email{pia.cortes@ug.uchile.cl}
         \and
             Department of Astronomy, Yale University, 
             New Haven, CT 06511, USA
         \and
            Chungnam National University, Department of Astronomy and Space Science,
            34134 Daejeon, Republic of Korea
        \and
            Aix Marseille Univ, CNRS, CNES, LAM,
            Marseille, France
        \and
            \textit{LSSTC DSFP} Fellow
        \and
            \textit{51 Pegasi b} Fellow
             }

   \date{Received July 11, 2019; accepted January 29, 2020 }

 
  \abstract
{We present 22 new transit observations of the exoplanets WASP-18Ab, WASP-19b, and WASP-77Ab, from the Transit Monitoring in the South (\emph{TraMoS}) project. We simultaneously model our newly collected transit light curves with archival photometry and radial velocity data, to obtain refined physical and orbital parameters. We include TESS light curves of the three exoplanets to perform an extended analysis of the variations in their transit mid-time (TTV) and to refine their planetary orbital ephemeris. We did not find significant $\rm TTV_{RMS}$ variations larger than 47, 65, and 86 seconds for WASP-18Ab, WASP-19b, and WASP-77Ab, respectively. Dynamical simulations were carried out to constrain the masses of a possible perturber. The observed RMS could be produced by a perturber body with an upper limit mass of 9, 2.5, 11 and $4~M_{\oplus}$ in 1:2, 1:3, 2:1, and 3:1 resonances in the WASP-18Ab system. In the case of WASP-19b, companions with masses up to 0.26, 0.65, 1 and $2.8~M_{\oplus}$, in 1:2, 2:1, 3:1, and 5:3 resonances, produce the RMS. And for the WASP-77Ab system, a planet with mass between $1.5-9~M_{\oplus}$ in 1:2, 1:3, 2:1, 2:3, 3:1, 3:5, 5:3 resonances. Comparing our results with RV variations, we discard massive companions with $350~M_{\oplus}$ in 17:5 resonance for WASP-18Ab, $95~M_{\oplus}$ in 4:1 resonance for WASP-19b and $105~M_{\oplus}$ in 5:2 resonance for WASP-77Ab. Finally, using a Lomb-Scargle period search we find no evidence of a periodic trend on our TTV data for the three exoplanets.}

   \keywords{}

   \maketitle
%

\section{Introduction}\label{intro}

High-precision long-term transit follow-ups provide tremendous opportunities in improving our understanding of exoplanets, leading to obtain more accurate measurements of planetary radius, especially those detected with ground-based transit surveys (e.g., HATNet and HATSouth, \citealt{Bakos2012}; SuperWASP, \citealt{Pollacco2006}; KELT, \citealt{Pepper2007}; TRES, \citealt{Alonso2007}, CSTAR, \citealt{WangS2014}). With improved photometry, we can refine planetary orbital ephemeris \citep{TEMP1}, which is vital to schedule future transit-related observations, such as Rossiter-Mclaughlin effect measurement \citep{Nutzman2011,Sanchis2011,Sanchis2013,WangS2018} and transmission spectrum follow-up \citep{Mancini2016b,Mackebrandt2017}.

Long-term photometric follow-up also provides a unique chance to study the variations of the orbital periods. A recent study shows the apparent orbital decay in the WASP-12 system \citep{Patra2017}, which intrigues a series of theoretical studies \citep{Millholland2018,Weinberg2017} to discuss the potential mechanisms. The transit follow-up also plays an important role in exoplanet system which shows interesting Transit Timing Variations (TTV) \citep{Ballard2011,Ford2012a,Steffen2012,Fabrycky2012,Mancini2016,WangS2017,Wu2018}. 
\cite{Ballard2018} predicted that around 5\% of planets discovered by TESS \citep{Ricker2014} will show TTVs. Transit follow-up of these targets is very critical because most of them will only be monitored for $\sim27$ days, whereas the typical TTV period is around years. 

Furthermore, extended TTV studies are crucial to confirm or rule out exoplanetary systems, in cases where space-based observations will not cover the long-time scales required to characterize them \citep{vonEssen2018}. Thus, combining ground and space-based observations will be crucial. 

The TTV method \citep{Miralda2002,Agol2005,Holman2005}  also provides a powerful tool to detect additional low-mass planets in hot Jupiter systems, which is usually hard to find by using other techniques \citep{Steffen2012b}. Many efforts have been devoted to this field \citep{Pal2011,Hoyer2012,Hoyer2013,Szabo2013}, but so far only two hot Jupiters have been found to accompanies with additional close-in planets (WASP-47: \cite{Becker2015}, and Kepler-730: \cite{Canas2019}). The accurate occurrence rate of the ``WASP-47-like" system, which hosts a hot Jupiter and at least one additional planet with a period less than 100 days, is still unknown.

To refine orbital parameters of currently known exoplanets, and to search for additional planets by using the TTV method, we organized the Transit Monitoring in the South hemisphere (TraMoS) project \citep{Hoyer2011} since 2008. We use one-meter class telescopes in the north of Chile to conduct high-precision long-term transit follow-up. 

Following the previous efforts from the TraMoS project, in this work, we present new light curves of three hot Jupiters: WASP-18Ab, WASP-19b, WASP-77Ab. Combining our new light curves, and archival photometric and radial velocity data sets, we refined the orbital and physical parameters of the systems and constrained the upper mass limit of potential additional planetary companions. 

WASP-18Ab is a transiting hot Jupiter discovered by \citet{Hellier2009} within the WASP-South transit survey \citep{Pollacco2006}. It is an extremely close-in planet with an orbital period of 0.94 days. The host star is an F6 type and it is the brightest component of a binary system \citep{Csizmadia2019,Fontanive2019}. Regarding its physical properties, WASP-18Ab is about ten times more massive than Jupiter with approximately the same radius ($M_{P}=10.3\,{\rm M_{Jupiter}}$, $R_{P}=1.1\,{\rm R_{Jupiter}}$). Even though a rapid orbital decay was predicted theoretically \citep{Hellier2009}, it is not observed yet \citep{Wilkins2017} and new theoretical models propose a variation of fewer than 4 seconds in the transit time over a 20-yr baseline
\citep{CollierCameron2018}.

The hot Jupiter WASP-19b was first reported by \cite{Hebb2010}. It is known as one of the hot Jupiters with the shortest orbital period ($P=0.788\,{\rm days}$). With a mass of $1.15\,{\rm M_{Jupiter}}$ and a radius of $1.31\,{\rm R_{Jupiter}}$, the planet orbits an active G8 dwarf.

The third exoplanet we followed-up in this work, WASP-77Ab, was first presented by \cite{Maxted2013}. WASP-77Ab has a mass of $1.8\,{\rm M_{Jupiter}}$ and a radius of $1.2\,{\rm R_{Jupiter}}$. It transits a G8 star in 1.36 days, which is the brightest component of a visual binary system. This system has a separation of 3.3 arcsec.

This paper is organized as follows. In Section~\ref{obs} are summarized the photometric observations and their reduction process. In Section~\ref{lc} we present the new light curves of the targets and the description of the technique used to obtain their orbital and physical parameters. The principal results and their consequences are presented in Section~\ref{res}. Finally, a summary and conclusions are described in Section~\ref{summary}.


\section{Observations and Data Reduction}\label{obs} 

\begin{table*}
\caption{Log of Observations}             
\label{log_table}      
\centering          
\begin{tabular}{c  c  c  c  c  c  c  c c c}
\hline\hline       
Target & Date & Epoch\tablefootmark{a} & Telescope & Filter & N\tablefootmark{b}& $\rm T_{exp}$\tablefootmark{c} & airmass & FWHM & RMS\tablefootmark{d} \\
      & (UTC) &       &           &       &                   & (sec) & & (arcsec) & (mmag)\\
\hline  
WASP-18 & 2009 Oct 28 &-1904 & SMARTS 1 m & $I$ & 1412 & 1.5 & $1.3\rightarrow1.0\rightarrow1.7$ & 1.18 & 8.49  \\
        & 2009 Oct 29 & -1903 & SMARTS 1 m & $I$ & 1435 &  2 & $1.4\rightarrow1.0\rightarrow1.6$ & 1.59 & 5.67 \\
        & 2009 Oct 30 & -1902 & SMARTS 1 m & $I$ & 1198 &  2 & $1.2\rightarrow1.0\rightarrow1.6$ & 1.93 & 4.50 \\
        & 2011 Sep 06 & -1184 & SMARTS 1 m & $I$ & 203 & 15 & $2.1\rightarrow1.6\rightarrow1.3$  & 5.66 &2.40 \\
        & 2016 Sep 24\tablefootmark{e} & 776 & Danish 1.54 m & $I$ & 138  & 90 & $1.0\rightarrow1.2\rightarrow1.5$ & 16.22 & 1.05  \\
        & 2016 Sep 25\tablefootmark{e} & 777 &Danish 1.54 m & $I$ & 159 & 90 & $1.0\rightarrow1.2\rightarrow1.5$& 17.55 & 0.96  \\
        & 2016 Sep 26\tablefootmark{e} & 778 & Danish 1.54 m & $I$ & 113 &  90 & $1.2\rightarrow1.0\rightarrow1.1$ & 17.94 & 0.87  \\ \smallskip
        & 2017 Sep 29\tablefootmark{e} & 1169 & Danish 1.54 m & $R$ & 330 & 30 & $1.0\rightarrow1.3\rightarrow1.5$ & 18.60 & 2.53 \\

WASP-19 & 2011 Apr 22 & -923 & SMARTS 1 m    & $I$ &  626 &  12 & $1.0\rightarrow1.4\rightarrow1.9$ & 0.72 & 4.31   \\
        & 2011 Dec 24 & -611 & SMARTS 1 m    & $I$ & 364  &  18 & $1.6\rightarrow1.3\rightarrow1.1$ & 1.42 & 35.9 \\
        & 2013 Mar 13 & -47  & Danish 1.54 m & $R$ & 336  & 35  & $1.3\rightarrow1.1\rightarrow1.2$ & 6.82 & 2.15\\
        & 2013 Apr 20 & 1    & Danish 1.54 m & $R$ & 153  & 100 & $1.1\rightarrow1.0\rightarrow1.3$ & 8.46 & 0.80 \\
        & 2015 Mar 04 & 867  & Danish 1.54 m & $R$ & 235  & 60  & $1.3\rightarrow1.0\rightarrow1.2$ & 3.99 & 0.84\\
        & 2016 Apr 14 & 1383 & Danish 1.54 m & $I$ & 87   & 100 & $1.0\rightarrow1.3\rightarrow1.6$ & 4.56 & 0.71\\
        & 2017 Feb 14 & 1771 & Danish 1.54 m & $I$ & 137  & 90  & $1.1\rightarrow1.0\rightarrow1.3$ & 4.21 & 0.79\\
        & 2017 Apr 08 & 1838 & Danish 1.54 m & $R$ & 125  & 90  & $1.1\rightarrow1.0\rightarrow1.2$ & 3.87 & 0.81\\\smallskip
        & 2017 Oct 03 & 2064 & Danish 1.54 m & $R$ & 43   & 110 & $3.0\rightarrow2.4\rightarrow1.8$ & 3.08 & 1.70\\ 

WASP-77 & 2013 Aug 20 & -659 & ETD\tablefootmark{f} & $clear$ & 103 & 120 & $2.5\rightarrow2.0\rightarrow1.4$ & --\tablefootmark{g}& 3.87  \\
        & 2013 Oct 30 & -606 & ETD\tablefootmark{f} & $clear$ & 690 &  12 & $1.7\rightarrow1.5\rightarrow1.6$ & --\tablefootmark{g}& 5.91 \\
        & 2015 Sep 29 & -92 & Danish 1.54 m         & $R$     & 244 &  30 & $1.2\rightarrow1.1\rightarrow1.3$ & 10.99 & 0.84\\
        & 2015 Oct 03 & -89 & Danish 1.54 m         & $R$     & 138 & 60  & $1.1\rightarrow1.4\rightarrow1.7$ & 12.09 & 1.84\\
        & 2016 Sep 26 & 175 & Danish 1.54 m         & $I$     & 90  & 90  & $1.1\rightarrow1.3\rightarrow1.6$ & 19.50 & 0.47\\
        & 2016 Sep 30 & 177 & ETD\tablefootmark{f}  & $clear$ & 66  & 180 & $1.5\rightarrow1.4\rightarrow1.8$ & --\tablefootmark{g} & 2.74 \\
        & 2016 Oct 07 & 183 & Warsaw 1.3 m          & $I$     & 237 &  60 & $1.4\rightarrow1.1\rightarrow1.3$ & 8.32 & 2.38 \\ 
        & 2016 Dec 09 & 229 & ETD\tablefootmark{f}  & $R$     & 57  & 180 & $2.4\rightarrow2.0\rightarrow1.4$ & 7.50 & 2.11 \\ 
        & 2017 Oct 01 & 447 & Danish 1.54 m         & $B$     & 224 & 30  & $1.2\rightarrow1.1\rightarrow1.4$ & 2.06  & 3.48\\
\hline                  
\end{tabular}
\tablefoot{
\tablefoottext{a}{The epoch 0 is $T_{0}$ in Tables~\ref{wasp18},~\ref{tab:wasp19} and \ref{tab:wasp77}, for WASP-18Ab, WASP-19b and WASP-77Ab, respectively.}
\tablefoottext{b}{Number of observations.}
\tablefoottext{c}{Exposure time of each observation. For the variable exposure times, we consider the average during the night.}
\tablefoottext{d}{The RMS values were computed from the best fitted model of each light curve.}
\tablefoottext{e}{Light curves computed with only one reference star.}
\tablefoottext{f}{Light curves obtained from the Exoplanet Transit Database (ETD) (\url{https://var2.astro.cz/ETD}).}
\tablefoottext{g}{Information not provided.}}
\end{table*}

We collected 8 light curves for WASP-18Ab between 2009 and 2017, 9 light curves for WASP-19b between 2011 and 2017, and 5 light curves for WASP-77Ab between 2015 and 2017. We included 4 transits of WASP-77Ab from the Exoplanet Transit Database (ETD: \cite{Poddan2010}) to cover a larger time span.

All the photometry was collected by using either the Danish 1.54 m telescope at ESO La Silla Observatory or the SMARTS 1 m at Cerro Tololo Observatory (CTIO), except for one transit of WASP-77Ab that was observed with the Warsaw 1.3 m at Las Campanas Observatory (LCO). The log of our observations is in Table \ref{log_table}. The new TraMoS light curves used for this work are presented in Figure~\ref{transits}.

For the photometric observations conducted on the Danish telescope, we used the Danish Faint Object Spectrograph and Camera (DFOSC) instrument, which has a $2{\rm K} \times 2{\rm K}$ CCD with a 13.7 x 13.7 arcmin$^2$ field of view (FoV) and a pixel scale of 0.39" per pixel. To reduce the readout time, some of the Danish 1.54 m images were windowed to only include the target star and its closest reference stars. The observations of the transits of WASP-18Ab during 2016 and 2017 were forced to be windowed due to a malfunction of the CCD. For those transits, only one reference star was used to perform the photometry.

The SMARTS 1 m has the Y4KCam instrument which is a $4{\rm K} \times 4{\rm K}$ CCD camera with a $20\times20$ arcmin$^2$ FoV and a pixel scale of 0.289" per pixel. 

For the observation with the Warsaw 1.3 m telescope, we used a $2048 \times 4096$ CCD camera chip with a 1.4 square degrees of FoV and 0.26" per pixel scale. No windowing or binning was used during the observations on both SMARTS 1m and the Warsaw 1.3m telescope.

As suggested by \cite{Southworth2009}, most of our observations, specifically those conducted after 2011, used the defocus technique. This allows longer exposure times in bright targets and improves the photometric precision. We adjusted the exposure time during the observations if the weather was not ideal. The recorded Julian Date in the Coordinated Universal Time (${\rm JD_{UTC}}$) were converted into Barycentric Julian Date in the Barycentric Dynamical Time standard (${\rm BJD_{TDB}}$) by following the procedure as in \citet{Eastman2010}.

We reduced the data by using our custom pipeline. It follows the standard procedures of reduction, calibration, and aperture photometry, but customized for each used instrument. The data was bias and flat-field calibrated using master bias and master flat-field images. These master images were built from at least 10 individual bias and flats obtained at the beginning of each observing night. When that was not possible, we used bias and flat-field images from the closest observing night. Then, the flux time series of all the selected stars in the FoV was obtained using aperture photometry. The radius of the aperture was chosen, in order to minimize the dispersion of the light curve in the out-of-transit points. To remove the sky-background we used the median of the pixels in a ring around the star. The size of this ring depends on each case, but it was determined in an iterative process over a range of values for external and internal sky radius. To built the relative photometry of the target we use the best reference stars in terms of its variability after checking for saturation and stability. The pipeline semi-automatically finds the best aperture and the size of the ring for the sky that produces the light curve with less RMS. 

\begin{table}
\caption{Example photometry of WASP-18A, WASP-19 and WASP-77A}
\label{examplephot}
\begin{tabular}{c c c c}
\hline \hline
Target & ${\rm BJD_{TDB}}$\tablefootmark{a} & Relative flux & Error\\
\hline
    WASP-18A &  2457658.658241   & 1.00168 & 0.00078 \\
             &  2457658.660591   & 1.00138 & 0.00080 \\
             &  2457658.661771   & 1.00195 & 0.00082 \\
             &  2457658.662940   & 1.00261 & 0.00085 \\
             &  2457658.664109   & 1.00137 & 0.00086 \\\smallskip
      & ...           & ...         &    ... \\
    WASP-19 & 2457086.543926   & 1.00099 & 0.00086  \\
             & 2457086.544916   & 1.00173 & 0.00091  \\
             & 2457086.545905   & 1.00139 & 0.00086  \\
             & 2457086.546895   & 1.00045 & 0.00094  \\
             & 2457086.547886   & 1.00064 & 0.00093  \\\smallskip
     &  ...           & ...         &    ...  \\
WASP-77A & 2457299.78624 & 1.00229 & 0.00028 \\ 
         & 2457299.78764 & 1.00116 & 0.00022 \\
         & 2457299.78855 & 1.00201 & 0.00022 \\
         & 2457299.78946 & 1.00216 & 0.00022 \\
         & 2457299.79092 & 1.00133 & 0.00021 \\
         & ...           & ...         &    ... \\    
\hline
\end{tabular}
\tablefoot{These tables are available in machine-readable form.\\
\tablefoottext{a}{The column time was converted to (${\rm BJD_{TDB}}$), following the procedure of \cite{Eastman2010}.}
}
\end{table}

\section{Light curve and RV analysis}\label{lc}

\begin{figure*}
\centering
\includegraphics[width=1.0\textwidth]{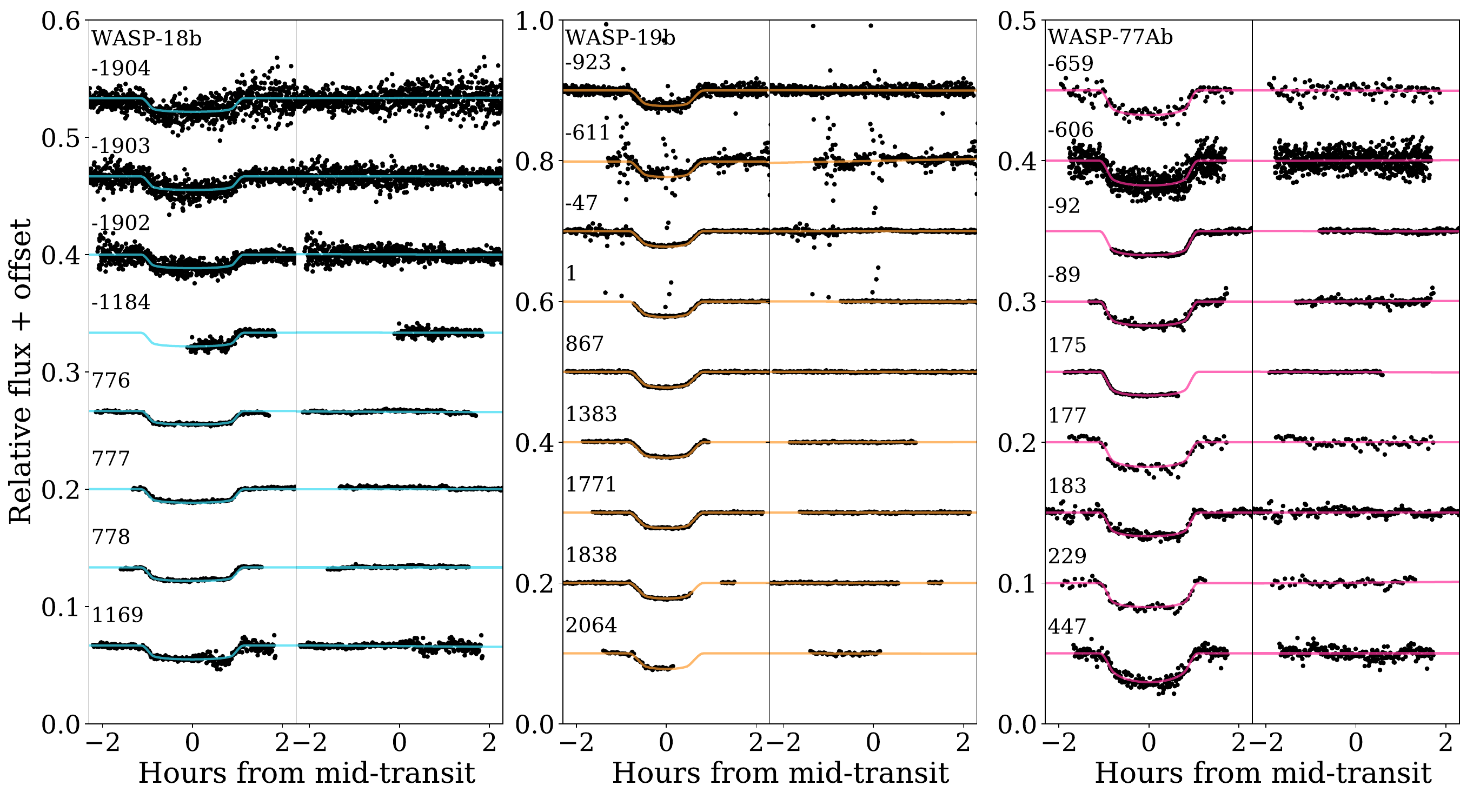}
\caption{Light curves from the TraMoS project of WASP-18A, WASP19 and, WASP77A during 8, 9 and 9 different transits, respectively. The best-fitted model from EXOFASTv2 is shown as a light blue solid line for WASP-18Ab, orange for WASP-19b and pink for WASP-77Ab. On the right of each panel are the corresponding residuals of the model. For clarity, both light curves and their residuals are offset artificially. The epoch number is indicated above each light curve. The technical information about each observation is listed in Table~\ref{log_table}.}
\label{transits}
\end{figure*}

\begin{figure*}
\includegraphics[width=1.0\textwidth]{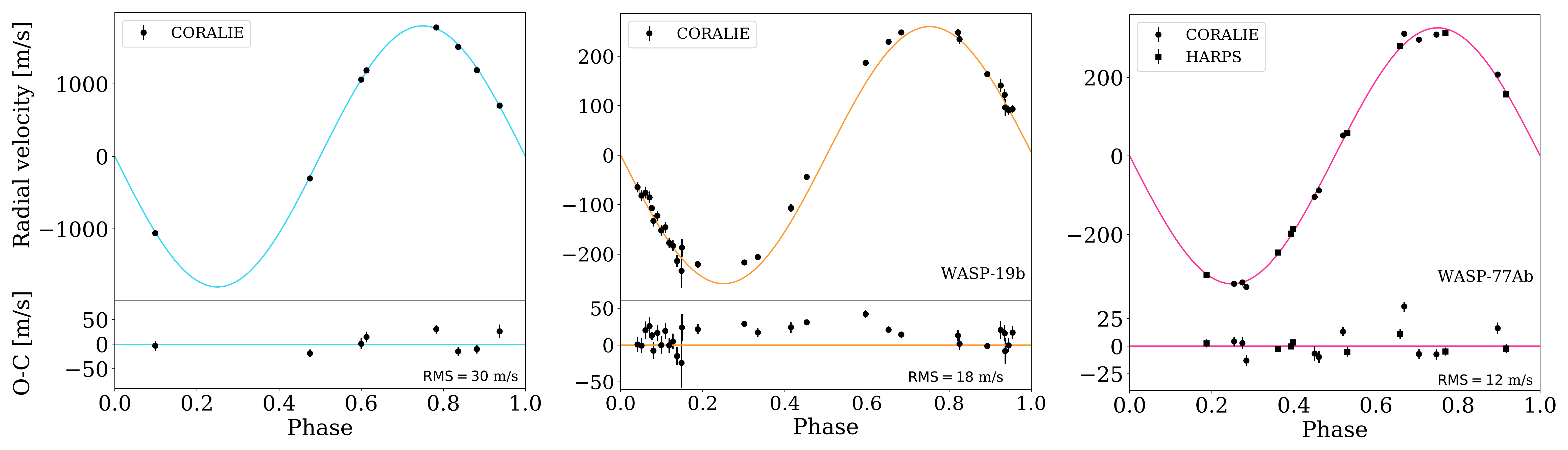}
\caption{Radial velocity observations of WASP-18A, WASP-19 and WASP-77A from \cite{Hellier2009}, \cite{Hebb2010} and \cite{Maxted2013}, respectively. The best-fitted model from the joint modeling of RV and light curves with EXOFASTv2 is in solid line color: light blue for WASP-18Ab, orange for WASP-19b and pink for WASP-77Ab. The residuals of the model are shown at the bottom panel of each figure.}
\label{rv}
\end{figure*}

To obtain the refined orbital and physical parameters of WASP-18Ab, WASP-19b, and WASP-77Ab, as well as their transit mid-time ($T_{c}$), we used EXOFASTv2 \citep{Eastman2013,Eastman2017} to model the light curves together with archived RV data from \cite{Hellier2009}, \cite{Hebb2010}, and \cite{Maxted2013}.

EXOFASTv2 is an IDL code designed to simultaneously fit transits and radial velocity measurements obtained from different filters or different telescopes. It uses the Differential Evolution Markov chain Monte Carlo (DE-MCMC) method to derive the values and their uncertainties of the stellar, orbital and physical parameters of the system. 

The stellar parameters of WASP-18A, WASP-19, and WASP-77A were computed using the MESA Isochrones and Stellar Tracks (MIST) model \citep{Dotter2016} included in EXOFASTv2. We applied Gaussian priors in surface gravity $\log{g}$, effective temperature $T_{\rm eff}$, and metallicity [Fe/H] of the stars, from \cite{Hellier2009}, \cite{Hebb2010} and \cite{Maxted2013} for WASP-18A, WASP-19 and WASP-77A, respectively. These priors have mixed origins. While the priors used in WASP-18A came from stellar evolutionary tracks models, for WASP-19 and WASP-77A their priors have a spectroscopic origin.

We were not able to separate the contribution of the two companions of the binary system WASP-77. The separation of the components is 3.3 arcsec, but our photometry aperture is about 10 arcsec. Thus, we computed the dilution factor -- fraction of the light that comes from the companion star -- for each filter of our data set to get the real transit depth of WASP-77Ab. Because of the lack of good quality magnitude measurements for the fainter companion WASP-77B in the $B$, $I$, $R$ and $clear$ passbands, we derived them from the \emph{Gaia} magnitude ($G=11.8356$) assuming Black Body radiation. The derived magnitudes for WASP-77B are $V=11.97$, $B=12.72$, $R=11.57$, $I=10.95$ and $\emph{clear}=11.78$.

We set previously published values as uniform priors for the DE-MCMC in all the transit, RV parameters, quadratic limb darkening coefficients and $T_{c}$. The priors were taken from the discovery papers of WASP-18Ab \citep{Hellier2009}, WASP-19b \citep{Hebb2010} and WASP-77Ab \citep{Maxted2013}.

In order to reduce significantly the convergence time of the chains during the EXOFASTv2 fitting, we started from shorter chains. Thus, the total time to complete that run is reduced. After it finished, we took the values from its best model and used them as priors for the next short run. This process was repeated until the chains were converged and well-mixed.

The best-fitted model is presented in Figure~\ref{transits} for our transit data from the TraMoS project, and in Figure~\ref{rv} for the RV archival data.

\begin{figure}
\centering
\includegraphics[width=1.0\columnwidth]{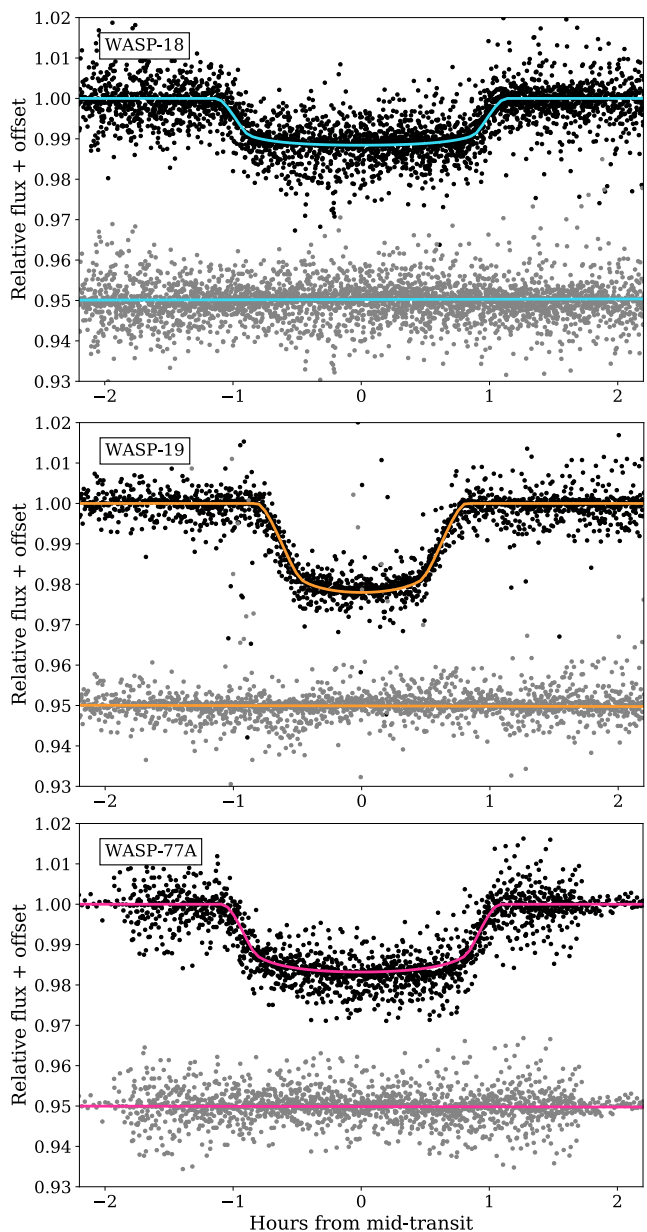}
\caption{Phased light curves of WASP-18Ab, WASP-19b and WASP-77Ab transits. The three data set of light curves are fitted simultaneously with RV archival data using EXOFASTv2, in order to estimate the orbital and physical parameters of the system. At the top panel, the light blue solid line is the best fitted model for WASP-18Ab, and bellow are the residuals in color grey. The same for WASP-19b in color orange (\emph{center panel}), and for WASP-77Ab in color pink (\emph{bottom panel}).}
\label{phase}
\end{figure}

\section{Results and Discussion}\label{res}

\subsection{Transit Parameters and Physical Properties}\label{transitparams}

\subsubsection{WASP-18Ab}

The resulting parameters from the global fit of the WASP-18A system in comparison with the results of the discovery paper \cite{Hellier2009} and the most recent analysis with TESS data \citep{Shporer2018}, are listed in Table~\ref{wasp18}. While in \cite{Hellier2009} the analysis was performed combining photometry and RV data, in \cite{Shporer2018} only photometric data was used. 

As the stellar spectroscopic priors were taken from the discovery paper \cite{Hellier2009}, our results for the stellar mass $M_*$ and radius $R_*$ are in good agreement with theirs, as expected, as well as the rest of the stellar parameters. \cite{Shporer2018} does not present results of stellar parameters.

In the case of the primary transit parameters, the greatest difference is found in the radius of the planet in stellar radii $R_{p}/R_{*}$. Our reported $R_{p}/R_{*}$ is $7.5\sigma$ and $4.1\sigma$ larger that the reported by \cite{Hellier2009} on the discovery paper and the recent result from \cite{Shporer2018}, respectively. Our transit duration $T_{14}$ is also $3.8\sigma$ larger than the value from \cite{Hellier2009}.

For the radial velocity parameters, the RV semi-amplitude derived from our analysis is consistent with the value of \cite{Hellier2009}, as the same data was used. 

Finally, the derived parameters of the system are, in general, in good agreement with the values from \cite{Hellier2009} and \cite{Shporer2018}. Even though our value for the eccentricity $e$ is within $1\sigma$ to the result from \cite{Hellier2009}, it is important to highlight that their difference may be a consequence as our limited number of RV measurement. We did not consider one RV measurement from \cite{Hellier2009} that were observed during a transit event.

\begin{table*}
\caption{System parameter of WASP-18A}
\label{wasp18}
\centering
\begin{tabular}{lcccc}
\hline \hline
~~~~Parameter & Units & This work & \cite{Hellier2009} & \cite{Shporer2018} \\
\hline
\\\multicolumn{2}{l}{Stellar Parameters:}&\smallskip\\
~~~~$M_*$\dotfill &Mass (\(M_\odot\))\dotfill &$1.294^{+0.063}_{-0.061}$ & $1.25\pm0.13$ &  \\
~~~~$R_*$\dotfill &Radius (\(R_\odot\))\dotfill &$1.319^{+0.061}_{-0.062}$ &$1.216^{+0.067}_{-0.054}$ & \\
~~~~$L_*$\dotfill &Luminosity (\(L_\odot\))\dotfill &$2.68^{+0.28}_{-0.26}$ & &\\
~~~~$\rho_*$\dotfill &Density (cgs)\dotfill &$0.795^{+0.11}_{-0.089}$ & $ 0.707^{+0.056}_{-0.096}$&\\
~~~~$\log{g}$\dotfill & Surface gravity (cgs)\dotfill &$4.310^{+0.036}_{-0.033}$ &$4.367^{+0.028}_{-0.042}$ & \\
~~~~$T_{\rm eff}$\dotfill &Effective Temperature (K)\dotfill &$6432\pm48$ & $6400\pm100$& \\
~~~~$[{\rm Fe/H}]$\dotfill &Metallicity \dotfill &$0.107\pm0.080$ & $0.00\pm0.09$ & \\
~~~~$Age$\dotfill &Age (Gyr)\dotfill &$1.57^{+1.4}_{-0.94}$ & $0.5-1.5$ & \\

\\\multicolumn{2}{l}{Planetary Parameters:}&\smallskip\\
~~~~$R_P$\dotfill &Radius (\rj)\dotfill &$1.240\pm0.079$ & $1.106^{+0.072}_{-0.054}$& $1.192\pm0.038$\\
~~~~$M_P$\dotfill &Mass (\mj)\dotfill &$10.20\pm0.35$ & $10.30\pm0.69$ & \\
~~~~$P$\dotfill &Period (days)\dotfill &$0.94145223\pm(24)$ & $0.94145299\pm(87)$& $0.9414576^{(+34)}_{(-35)}$ \\
~~~~$e$\dotfill &Eccentricity \dotfill &$0.0051^{+0.0070}_{-0.0037}$ & $0.0092\pm0.0028$ &\\
~~~~$a$\dotfill &Semi-major axis (AU)\dotfill &$0.02024^{+0.00029}_{-0.00031}$ & $0.02045\pm0.00067$&\\
~~~~$\omega_*$\dotfill &Argument of Periastron (Degrees)\dotfill &$-85^{+72}_{-96}$ &  &\\
~~~~$\rho_P$\dotfill &Density (cgs)\dotfill &$6.6^{+1.2}_{-1.1}$& $7.73^{+0.78}_{-1.27}$\tablefootmark{b} & \\
~~~~$logg_P$\dotfill &Surface gravity \dotfill &$4.215^{+0.046}_{-0.052}$ & $4.289^{+0.027}_{-0.050}$ &\\
~~~~$T_{eq}$\dotfill &Equilibrium temperature (K)\dotfill &$2429^{+77}_{-70}$ & $2384^{+58}_{-30}$ & \\
~~~~$\Theta$\dotfill &Safronov Number \dotfill &$0.268^{+0.016}_{-0.017}$ & &\\
~~~~$\fave$\dotfill &Incident Flux (\fluxcgs)\dotfill &$7.90^{+1.10}_{-0.87}$ & & \\

\\\multicolumn{2}{l}{Primary Transit Parameters:}&\smallskip\\
~~~~$T_0$\dotfill &Transit time (\bjdtdb)\dotfill &$2456740.80560\pm(19)$ & $2454221.48163\pm(38)$ & $2458361.048072^{(+34)}_{(-35)}$\\
~~~~$i$\dotfill &Inclination (Degrees)\dotfill &$83.5^{+2.0}_{-1.6}$ & $86.0\pm2.5$ & $84.31^{+0.40}_{-0.37}$\\
~~~~$R_P/R_*$\dotfill &Radius of planet in stellar radii \dotfill &$0.1018\pm0.0011$ &  $0.0935\pm0.0011$ & $0.09721^{+0.00016}_{-0.00017}$\\
~~~~$a/R_*$\dotfill &Semi-major axis in stellar radii \dotfill &$3.48^{+0.16}_{-0.17}$ & & $3.523^{+0.028}_{-0.027}$\\
~~~~$b$\dotfill &Impact parameter \dotfill &$0.36^{+0.11}_{-0.18}$ & $0.25\pm0.15$ & $0.349^{+0.020}_{-0.022}$\\
~~~~$\delta$\dotfill &Transit depth (fraction)\dotfill &$0.01041\pm0.00022$ & & $0.009449^{+0.000032}_{-0.000032}$\\
~~~~$u_{1,I}$\dotfill &linear LD coeff., I band\dotfill &$0.207\pm0.019$ & &\\
~~~~$u_{2,I}$\dotfill &quadratic LD coeff., I band\dotfill &$0.313\pm0.019$& &\\
~~~~$u_{1,R}$\dotfill &linear LD coeff., R band\dotfill & $0.257\pm0.045$ & &\\
~~~~$u_{2,R}$\dotfill &quadratic LD coeff., R band\dotfill &$0.309\pm0.048$ & &\\
~~~~$T_{14}$\dotfill &Total transit duration (days)\dotfill &$0.0921^{+0.0013}_{-0.0011}$ & $0.08932\pm0.00068$ &\\
~~~~$P_T$\dotfill &A priori non-grazing transit prob \dotfill &$0.258^{+0.014}_{-0.011}$ & &\\
~~~~$P_{T,G}$\dotfill &A priori transit prob \dotfill &$0.316^{+0.017}_{-0.014}$ & &\\
~~~~$\tau$\dotfill &Ingress/egress transit duration (days)\dotfill &$0.0099\pm0.0012$ & &\\

\\\multicolumn{2}{l}{RV Parameters:}&\smallskip\\
~~~~$e\cos{\omega_*}$\dotfill & \dotfill &$0.0002^{+0.0033}_{-0.0028}$ & &\\
~~~~$e\sin{\omega_*}$\dotfill & \dotfill &$-0.0022^{+0.0039}_{-0.0079}$ & &\\
~~~~$K$\dotfill &RV semi-amplitude (m/s)\dotfill &$1814^{+23}_{-24}$ & $1818.3\pm8.0$ &\\
~~~~$M_P\sin i$\dotfill &Minimum mass (\mj)\dotfill &$10.14\pm0.33$ & &\\
\\\multicolumn{2}{l}{Secondary Eclipse Parameters:}&\smallskip\\
~~~~$T_S$\dotfill &Time of eclipse (\bjdtdb)\dotfill &$2457657.3119^{+0.0021}_{-0.0019}$ & &\\
~~~~$b_S$\dotfill &Eclipse impact parameter \dotfill &$0.35^{+0.11}_{-0.17}$ & &\\
~~~~$\tau_S$\dotfill &Ingress/egress eclipse duration (days)\dotfill &$0.0098^{+0.0013}_{-0.0010}$ & &\\
~~~~$T_{S,14}$\dotfill &Total eclipse duration (days)\dotfill &$0.0917\pm0.0016$ & &\\
~~~~$P_S$\dotfill &A priori non-grazing eclipse prob \dotfill &$0.259^{+0.013}_{-0.012}$ & &\\
~~~~$P_{S,G}$\dotfill &A priori eclipse prob \dotfill &$0.318^{+0.017}_{-0.015}$ & &\\
\hline
\end{tabular}
\tablefoot{
\tablefoottext{a}{Value converted to cgs units multiplying by the Sun density $\rho_{\odot}=1.408\,$cgs.}
\tablefoottext{b}{Value converted to cgs units multiplying by the Jupiter density $\rho_{J}=1.33\,$cgs.}
\tablefoottext{c}{Values enclosed in parentheses correspond to the uncertainties of the last digits of the nominal value.}}
\end{table*}

\subsubsection{WASP-19b}
The results of the global fit of the WASP-19 system are listed in Table~\ref{tab:wasp19}, in comparison with the previous values from the discovery paper \citep{Hebb2010}, and a more recent work \citep{Lendl2013}.

To estimate the stellar parameters of WASP-19, we used as priors the stellar spectroscopic parameters from \cite{Hebb2010}. Thus, in general, our results are in agreement with those from the discovery paper. The most important discrepancies are the density of the star $\rho_*$ and the surface gravity $\log{g}$, showing $+2.5\sigma$ and $-3.2\sigma$ difference, respectively. Comparing with the results from \cite{Lendl2013}, ours are all in good agreement. The stellar surface gravity $\log{g}$ derived from spectroscopy may be different from the values that include constraints from transit data \citep{Torres2012}.

For values of the primary transit parameters obtained from the light curves, the greatest differences are found in the orbital inclination $i$ and the total transit duration $T_{14}$. We report an inclination value $5.1\sigma$ smaller than \cite{Hebb2010}, but in agreement with the estimate of \cite{Lendl2013}. In the other hand, our estimation of $T_{14}$ is significantly larger than \cite{Hebb2010} by $9\sigma$, but the difference is only $3.5\sigma$ when compared with \cite{Lendl2013}. We also report a more precise impact parameter $b$ and transit depth $\delta$.

As the same RV data set from the discovery paper \citep{Hebb2010} was used to perform our analysis, the almost identical values in the RV semi-amplitude $K$ is not a surprise. Moreover, the values from \cite{Lendl2013} are also in agreement. 

The planetary parameters derived from the light curve and radial velocity analysis are almost all in good agreement with the comparison works. The only parameter with a difference greater than $3\sigma$ is our estimation of the Equilibrium Temperature $T_{eq}$ compared with the result of \cite{Hebb2010}. However, our result is in better agreement with \cite{Lendl2013} by less than $2\sigma$.

\begin{table*}
\caption{System parameter of WASP-19}
\label{tab:wasp19}
\centering
\begin{tabular}{lcccc}
\hline \hline
~~~Parameter & Units & This work & \cite{Hebb2010}\tablefootmark{a} & \cite{Lendl2013}\\
\hline
\\\multicolumn{2}{l}{Stellar Parameters:}&\\
~~~~$M_*$\dotfill &Mass (\(M_\odot\))\dotfill &$0.965^{+0.091}_{-0.095}$ & $0.95\pm0.10$ & $0.968^{+0.084}_{-0.079}$ \\
~~~~$R_*$\dotfill &Radius (\(R_\odot\))\dotfill &$1.006^{+0.031}_{-0.034}$ & $0.93^{+0.05}_{-0.04}$& $0.994\pm0.031$\\
~~~~$L_*$\dotfill &Luminosity (\(L_\odot\))\dotfill &$0.905^{+0.071}_{-0.069}$\\
~~~~$\rho_*$\dotfill &Density (cgs)\dotfill &$1.339^{+0.056}_{-0.058}$ & $1.19^{+0.12}_{-0.11}$\tablefootmark{b}  & $1.384^{+0.055}_{-0.051}$\tablefootmark{b}\\
~~~~$\log{g}$\dotfill &Surface gravity (cgs)\dotfill &$4.417^{+0.020}_{-0.021}$ & $4.48\pm0.03$\\
~~~~$T_{\rm eff}$\dotfill &Effective Temperature (K)\dotfill &$5616^{+66}_{-65}$ & $5500\pm100$\\
~~~~$[{\rm Fe/H}]$\dotfill &Metallicity \dotfill &$0.04^{+0.25}_{-0.30}$ & $0.02\pm0.09$\\
~~~~$Age$\dotfill &Age (Gyr)\dotfill &$6.4^{+4.1}_{-3.5}$ & $5.5^{+9.0}_{-4.5}$\\

\\\multicolumn{2}{l}{Planetary Parameters:}&\\
~~~~$R_P$\dotfill &Radius (\rj)\dotfill &$1.415^{+0.044}_{-0.048}$ & $1.28\pm0.07$ & $1.376\pm0.046$\\
~~~~$M_P$\dotfill &Mass (\mj)\dotfill &$1.154^{+0.078}_{-0.080}$ & $1.14\pm0.07$ & $1.165\pm0.068$\\
~~~~$P$\dotfill &Period (days)\dotfill &$0.78883852^{+(75)}_{-(82)}$ & $0.7888399\pm(8)$  & $0.7888390\pm(2)$\\
~~~~$e$\dotfill &Eccentricity \dotfill &$0.0126^{+0.014}_{-0.0089}$ & & $0.0077^{+0.0068}_{-0.0032}$\\
~~~~$a$\dotfill &Semi-major axis (AU)\dotfill &$0.01652^{+0.00050}_{-0.00056}$ & $0.0164^{+0.0005}_{-0.0006}$ & $0.01653\pm0.00046$\\
~~~~$\omega_*$\dotfill &Argument of Periastron (Degrees)\dotfill &$51^{+89}_{-190}$ & $-76^{+112}_{-23}$ & $43^{+28}_{-67}$\\
~~~~$\rho_P$\dotfill &Density (cgs)\dotfill &$0.506^{+0.031}_{-0.030}$ & $0.54^{+0.07}_{-0.06}$& $0.595^{+0.036}_{-0.033}$\tablefootmark{c}\\
~~~~$logg_P$\dotfill &Surface gravity \dotfill &$3.155^{+0.018}_{-0.019}$ & $3.20\pm0.03$ & $3.184\pm0.015$\\
~~~~$T_{eq}$\dotfill &Equilibrium temperature (K)\dotfill &$2113\pm29$ & $1993^{+32}_{-33}$ & $2058\pm40$\\
~~~~$\Theta$\dotfill &Safronov Number \dotfill &$0.0279^{+0.0012}_{-0.0011}$\\
~~~~$\fave$\dotfill &Incident Flux (\fluxcgs)\dotfill &$4.52^{+0.26}_{-0.24}$\\

\\\multicolumn{2}{l}{Primary Transit Parameters:}&\\
~~~~$T_0$\dotfill & Transit Time (\bjdtdb)\dotfill &$2456402.7128^{+(17)}_{-(14)}$ & $2454775.3372\pm(2)$ & $2456029.59204\pm(13)$\\
~~~~$i$\dotfill &Inclination (Degrees)\dotfill &$79.08^{+0.34}_{-0.37}$ & $80.8\pm0.8$  & $79.54\pm0.33$\\
~~~~$R_P/R_*$\dotfill &Radius of planet in stellar radii \dotfill &$0.14410^{+0.00049}_{-0.00050}$ & $0.1425\pm0.0014$\\
~~~~$a/R_*$\dotfill &Semi-major axis in stellar radii \dotfill &$3.533^{+0.048}_{-0.052}$ &  & $3.573\pm0.046$\\
~~~~$b$\dotfill &Impact parameter \dotfill &$0.6671^{+0.0087}_{-0.0091}$ & $0.62\pm0.03$ & $0.645\pm0.012$\\
~~~~$\delta$\dotfill &Transit depth (fraction)\dotfill &$0.02077\pm0.00014$ & $0.0203\pm0.0004$ & $0.02018\pm0.00021$\\
~~~~$u_{1,I}$\dotfill &linear LD coeff., I band\dotfill &$0.287^{+0.027}_{-0.029}$&\\
~~~~$u_{2,I}$\dotfill &quadratic LD coeff., I band\dotfill &$0.263\pm0.024$&\\
~~~~$u_{1,R}$\dotfill &linear LD coeff., R band \dotfill &$0.383^{+0.029}_{-0.032}$\\
~~~~$u_{2,R}$\dotfill &quadratic LD coeff., R band\dotfill &$0.246^{+0.027}_{-0.025}$\\
~~~~$T_{14}$\dotfill &Total transit duration (days)\dotfill &$0.06697^{+0.00031}_{-0.00030}$ & $0.0643^{+0.0006}_{-0.0007}$ & $0.06586^{+0.00033}_{-0.00031}$\\
~~~~$P_T$\dotfill &A priori non-grazing transit prob \dotfill &$0.2426^{+0.0066}_{-0.0051}$\\
~~~~$P_{T,G}$\dotfill &A priori transit prob \dotfill &$0.3246^{+0.0089}_{-0.0069}$\\
~~~~$\tau$\dotfill &Ingress/egress transit duration (days)\dotfill &$0.01459\pm0.00035$\\

\\\multicolumn{2}{l}{RV Parameters:}&\\
~~~~$e\cos{\omega_*}$\dotfill & \dotfill &$-0.0027^{+0.0077}_{-0.013}$ & $0.004\pm0.009$ &$0.0024\pm0.0020$ \\
~~~~$e\sin{\omega_*}$\dotfill & \dotfill &$0.0016^{+0.014}_{-0.0092}$ & $-0.02\pm0.02$ & $0.000\pm0.005$\\
~~~~$K$\dotfill &RV semi-amplitude (m/s)\dotfill &$255.4^{+6.1}_{-6.2}$ & $256\pm5$  & $257.7\pm2.9$\\
~~~~$M_P\sin i$\dotfill &Minimum mass (\mj)\dotfill &$1.133^{+0.078}_{-0.079}$\\

\\\multicolumn{2}{l}{Secondary Eclipse Parameters:}&\\
~~~~$T_S$\dotfill &Time of eclipse (\bjdtdb)\dotfill &$2455169.3621^{+(41)}_{-(51)}$ & $2456030.77766\pm(88)$\\
~~~~$b_S$\dotfill &Eclipse impact parameter \dotfill &$0.670^{+0.020}_{-0.017}$ & $0.652\pm0.015$\\
~~~~$\tau_S$\dotfill &Ingress/egress eclipse duration (days)\dotfill &$0.01472^{+0.00085}_{-0.00066}$\\
~~~~$T_{S,14}$\dotfill &Total eclipse duration (days)\dotfill &$0.06812^{+0.00087}_{-0.00074}$\\
~~~~$P_S$\dotfill &A priori non-grazing eclipse prob \dotfill &$0.2415\pm0.0021$\\
~~~~$P_{S,G}$\dotfill &A priori eclipse prob \dotfill &$0.3232\pm0.0030$\\
\hline
\end{tabular}
\tablefoot{
\tablefoottext{a}{For comparison, the results from \cite{Hellier2009} that considered free eccentricity were used.}
\tablefoottext{b}{Values converted to cgs units multiplying by the Sun density $\rho_{\odot}=1.408\,$cgs.}
\tablefoottext{c}{Values converted to cgs units multiplying by the Jupiter density $\rho_{J}=1.33\,$cgs}
\tablefoottext{d}{Values enclosed in parentheses correspond to the uncertainties of the last digits of the nominal value.}}
\end{table*}

\subsubsection{WASP-77Ab}
Table~\ref{tab:wasp77} lists the results of the global fit of the WASP-77A system, in comparison with the values from its discovery paper \citep{Maxted2013} on which photometry and RV data were used. No other previous work has reported bulk measurements for this system.

Almost all the stellar parameters are in agreement with \citep{Maxted2013}, except for a $-9.7\sigma$ difference in the stellar surface gravity $\log{g}$, where our reported value is more precise. This difference can be explained as the same for the case of WASP-19 (see Section 4.2.1).

The primary transit parameters, as well as the RV parameters and the derived planetary parameters, are consistent with the results from \cite{Maxted2013}.

\begin{table*}
\caption{System parameters of WASP-77A}
\label{tab:wasp77}
\centering
\begin{tabular}{lcccc}
\hline \hline
~~~~~Parameter & Units & This work & \cite{Maxted2013}\\
\hline
\multicolumn{2}{l}{Stellar Parameters:}&\\
~~~~$M_*$\dotfill &Mass (\(M_\odot\))\dotfill &$0.903^{+0.066}_{-0.059}$ & $1.002\pm0.045$\\
~~~~$R_*$\dotfill &Radius (\(R_\odot\))\dotfill &$0.910^{+0.025}_{-0.023}$ & $0.955\pm0.015$\\
~~~~$L_*$\dotfill &Luminosity (\(L_\odot\))\dotfill &$0.743^{+0.065}_{-0.058}$ & \\
~~~~$\rho_*$\dotfill &Density (cgs)\dotfill &$1.692^{+0.056}_{-0.069}$\tablefootmark{a} & $1.629^{+0.023}_{-0.028}$\tablefootmark{a}\\
~~~~$\log{g}$\dotfill &Surface gravity (cgs)\dotfill &$4.476^{+0.014}_{-0.015}$ & $4.33\pm0.08$\\
~~~~$T_{\rm eff}$\dotfill &Effective Temperature (K)\dotfill &$5617\pm72$ & $5500\pm80$\\
~~~~$[{\rm Fe/H}]$\dotfill &Metallicity \dotfill &$-0.10^{+0.10}_{-0.11}$ & $0.00\pm0.11$\\
~~~~$Age$\dotfill &Age (Gyr)\dotfill &$6.2^{+4.0}_{-3.5}$ & $0.5-1.0$\\

\multicolumn{2}{l}{Planetary Parameters:}&\\
~~~~$R_P$\dotfill &Radius (\rj)\dotfill &$1.230^{+0.031}_{-0.029}$ & $1.21\pm0.02$\\
~~~~$M_P$\dotfill &Mass (\mj)\dotfill &$1.667^{+0.068}_{-0.064}$ & $1.76\pm0.06$\\
~~~~$P$\dotfill &Period (days)\dotfill &$1.36002854\pm(62)$ & $1.3600309\pm(20)$\\
~~~~$e$\dotfill &Eccentricity \dotfill&$0.0074^{+0.0069}_{-0.0049}$\\
~~~~$a$\dotfill &Semi-major axis (AU)\dotfill &$0.02335^{+0.00045}_{-0.00043}$ & $0.0240\pm0.00036$\\
~~~~$\omega_*$\dotfill &Argument of Periastron (Degrees)\dotfill &$-166^{+66}_{-75}$\\
~~~~$\rho_P$\dotfill &Density (cgs)\dotfill &$1.115^{+0.052}_{-0.062}$ & $1.33\pm0.04$\tablefootmark{b}\\
~~~~$logg_P$\dotfill &Surface gravity \dotfill &$3.438^{+0.012}_{-0.016}$ & $3.441\pm0.008$\\
~~~~$T_{eq}$\dotfill &Equilibrium temperature (K)\dotfill &$1715^{+26}_{-25}$\\
~~~~$\Theta$\dotfill &Safronov Number \dotfill &$0.0689\pm0.0018$\\
~~~~$\fave$\dotfill &Incident Flux (\fluxcgs)\dotfill &$1.96^{+0.12}_{-0.11}$\\

\multicolumn{2}{l}{Primary Transit Parameters:}&\\
~~~~$T_0$\dotfill &Transit Time (\bjdtdb)\dotfill &$2457420.88439^{(+80)}_{(-85)}$ & $2455870.44977\pm(20)$\\
~~~~$i$\dotfill &Inclination (Degrees)\dotfill &$88.91^{+0.74}_{-0.95}$ & $89.4^{+0.4}_{-0.7}$\\
~~~~$R_P/R_*$\dotfill &Radius of planet in stellar radii \dotfill &$0.13354^{+0.00074}_{-0.00070}$\\
~~~~$a/R_*$\dotfill &Semi-major axis in stellar radii \dotfill &$5.332^{+0.057}_{-0.081}$\\
~~~~$b$\dotfill &Impact parameter \dotfill &$0.109^{+0.089}_{-0.071}$ & $0.06^{+0.07}_{-0.05}$\\
~~~~$\delta$\dotfill &Transit depth (fraction)\dotfill &$0.01783^{+0.00020}_{-0.00019}$\\
~~~~$u_{1,B}$\dotfill &linear LD coeff., B band \dotfill &$0.680\pm0.054$&\\
~~~~$u_{2,B}$\dotfill &quadratic LD coeff., B band\dotfill &$0.140^{+0.052}_{-0.053}$&\\
~~~~$u_{1,clear}$\dotfill &linear LD coeff., \emph{clear} band \dotfill &$0.386\pm0.029$\\
~~~~$u_{2,clear}$\dotfill &quadratic LD coeff., \emph{clear} band \dotfill &$0.227\pm0.029$&\\
~~~~$u_{1,I}$\dotfill &linear LD coeff., I band \dotfill &$0.311\pm0.025$&\\
~~~~$u_{2,I}$\dotfill &quadratic LD coeff., I band \dotfill &$0.294\pm0.033$&\\
~~~~$u_{1,R}$\dotfill &linear LD coeff., R band \dotfill &$0.312\pm0.023$\\
~~~~$u_{2,R}$\dotfill &quadratic LD coeff., R band \dotfill &$0.237^{+0.029}_{-0.028}$\\
~~~~$T_{14}$\dotfill &Total transit duration (days)\dotfill &$0.08952^{+0.00053}_{-0.00051}$\\
~~~~$P_T$\dotfill &A priori non-grazing transit prob \dotfill &$0.1578^{+0.0029}_{-0.0025}$\\
~~~~$P_{T,G}$\dotfill &A priori transit prob \dotfill &$0.2064^{+0.0039}_{-0.0033}$\\
~~~~$\tau$\dotfill &Ingress/egress transit duration (days)\dotfill &$0.01075^{+0.00032}_{-0.00015}$\\

\multicolumn{2}{l}{RV Parameters:}&\\
~~~~$e\cos{\omega_*}$\dotfill & \dotfill &$-0.0039^{+0.0041}_{-0.0051}$\\
~~~~$e\sin{\omega_*}$\dotfill & \dotfill &$-0.0003^{+0.0061}_{-0.0076}$\\
~~~~$K$\dotfill &RV semi-amplitude (m/s)\dotfill &$323.4^{+3.8}_{-3.4}$ & $321.9\pm3.9$\\
~~~~$M_P\sin i$\dotfill &Minimum mass (\mj)\dotfill &$1.667^{+0.068}_{-0.064}$\\

\multicolumn{2}{l}{Secondary Eclipse Parameters:}&\\
~~~~$T_S$\dotfill &Time of eclipse (\bjdtdb)\dotfill &$2457658.2054^{+0.0036}_{-0.0044}$\\
~~~~$b_S$\dotfill &Eclipse impact parameter \dotfill &$0.109^{+0.092}_{-0.081}$\\
~~~~$\tau_S$\dotfill &Ingress/egress eclipse duration (days)\dotfill &$0.01116^{+0.00041}_{-0.00025}$\\
~~~~$T_{S,14}$\dotfill &Total eclipse duration (days)\dotfill &$0.0922^{+0.0012}_{-0.0014}$\\
~~~~$P_S$\dotfill &A priori non-grazing eclipse prob \dotfill &$0.1624^{+0.0022}_{-0.0012}$\\
~~~~$P_{S,G}$\dotfill &A priori eclipse prob \dotfill &$0.2126^{+0.0031}_{-0.0015}$\\
\hline
\end{tabular}
\tablefoot{
\tablefoottext{a}{Value converted to cgs units multiplying by the Sun density $\rho_{\odot}=1.408\,$cgs.}
\tablefoottext{b}{Value converted to cgs units multiplying by the Jupiter density $\rho_{J}=1.33\,$cgs.}
\tablefoottext{c}{Values enclosed in parentheses correspond to the uncertainties of the last digits of the nominal value.}
}
\end{table*}

\subsection{Transit Timing Variations}\label{ttvsection}

A transit timing variation (TTV) is represented through a difference in time between the expected transit mid-time, assuming a Keplerian motion for the planet, and the observed transit mid-time. The TTVs for the three targets were computed considering our transit mid-times from the TraMoS project, as well as including previous transit mid-times already published and new transit mid-times coming from TESS \citep{Ricker2014} light curves.
 
During its first year, TESS observed stars exclusively in the Southern hemisphere. WASP-18A was observed during Sector 2 and 3 producing 45 complete transit events. WASP-19 was observed in Sector 9 producing 29 complete transit events, and WASP-77A was observed during Sector 4 and produced 15 complete light curves.
 
TESS data are reduced by the Science Processing Operations Center (SPOC) and after being processed, they are archived in the Mikulski Archive for Space Telescopes (MAST\footnote{\url{https://exo.mast.stsci.edu/}}) catalog where can be downloaded directly by anyone. We downloaded the complete light curves of our three targets from the MAST catalog. Then, the transit events were identified and cut in independent light curves. For each TESS light curve, its corresponding transit mid-time $T_c$ was computed using EXOFASTv2 \citep{Eastman2013}. The transit mid-times of all the new TESS light curves are listed in Tables~\ref{times_wasp18}, \ref{times_wasp19}, and \ref{times_wasp77}.  
 
\begin{figure*}
    \centering
        \includegraphics[width=.73\textwidth]{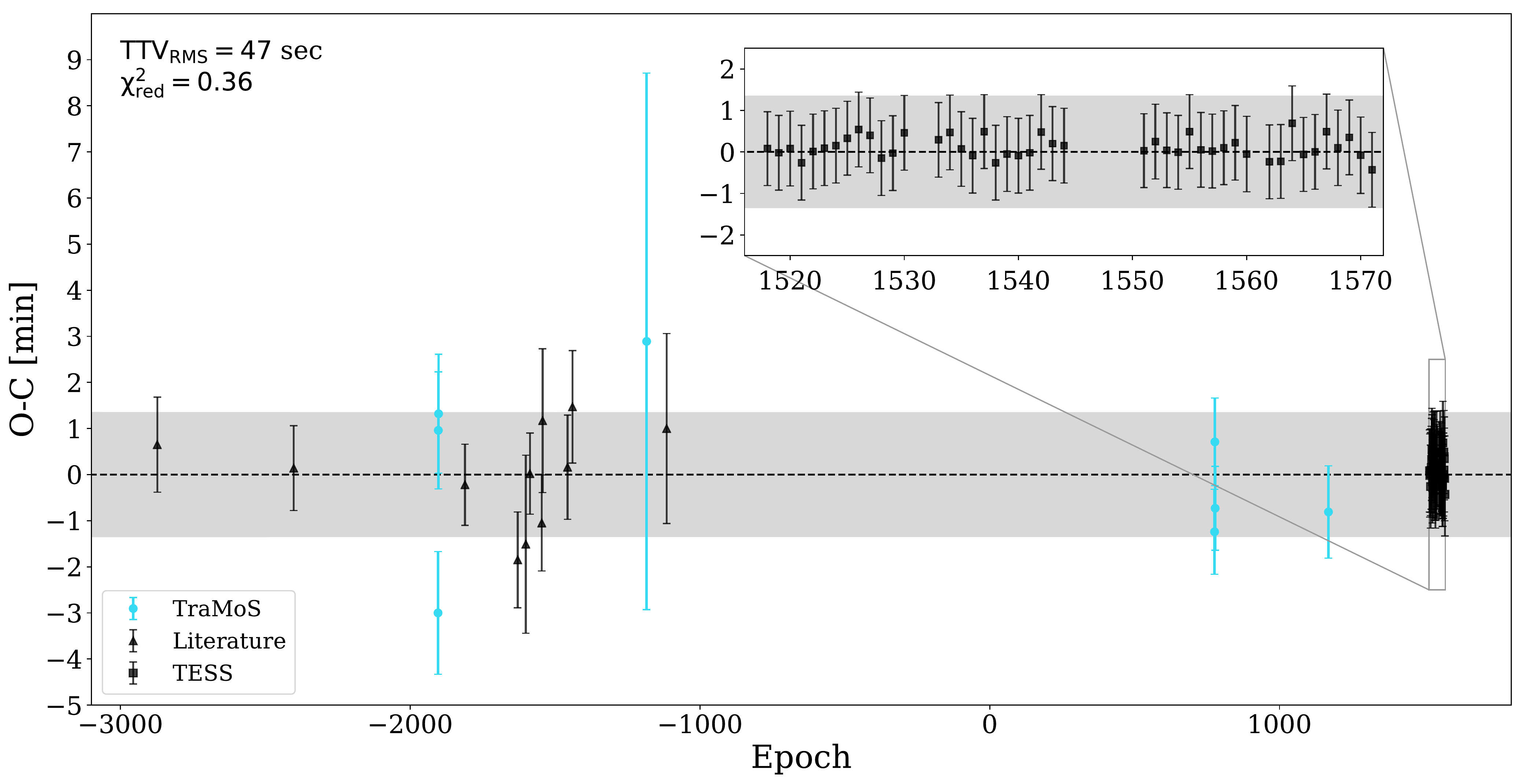}
        \includegraphics[width=.73\textwidth]{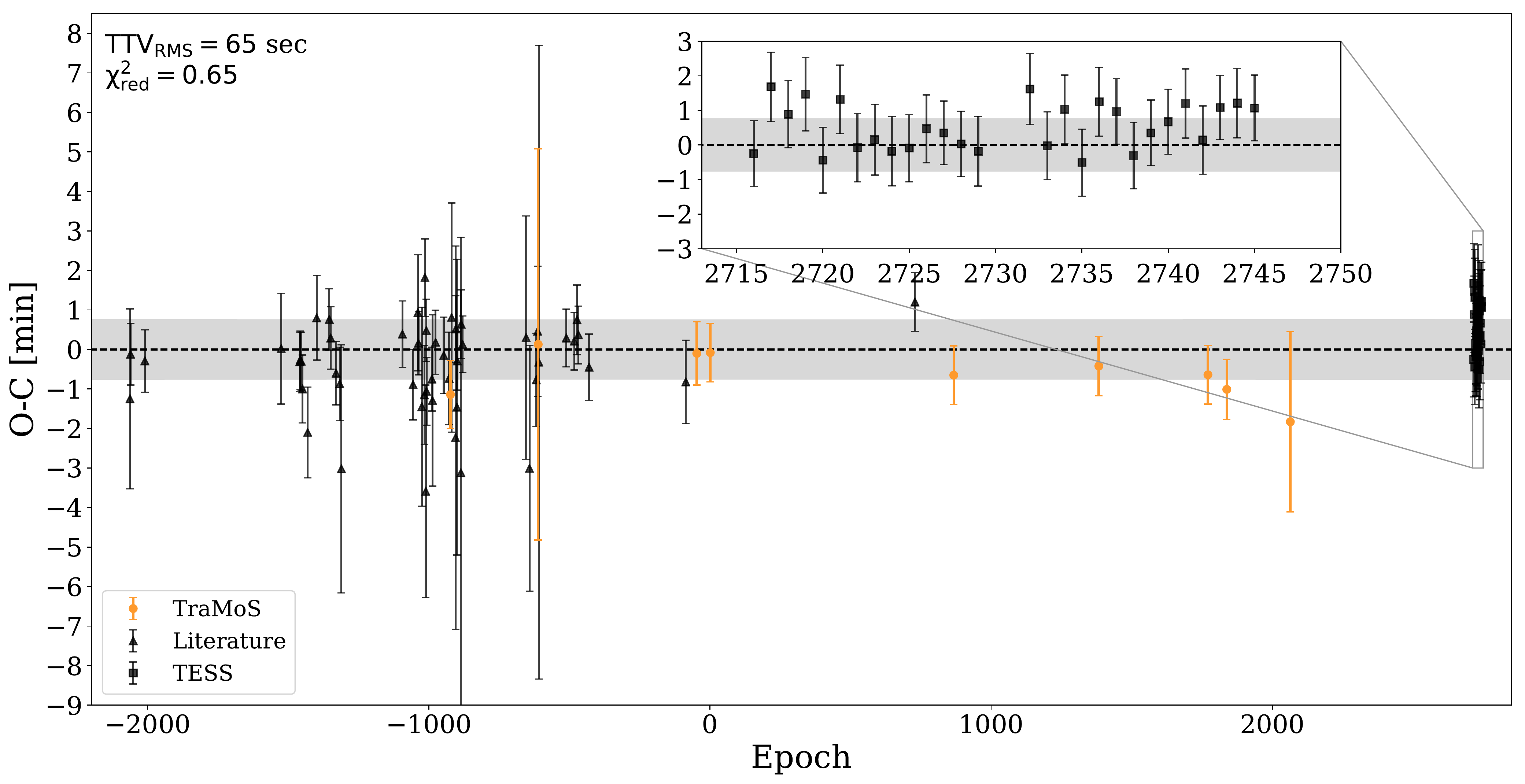}
        \includegraphics[width=.73\textwidth]{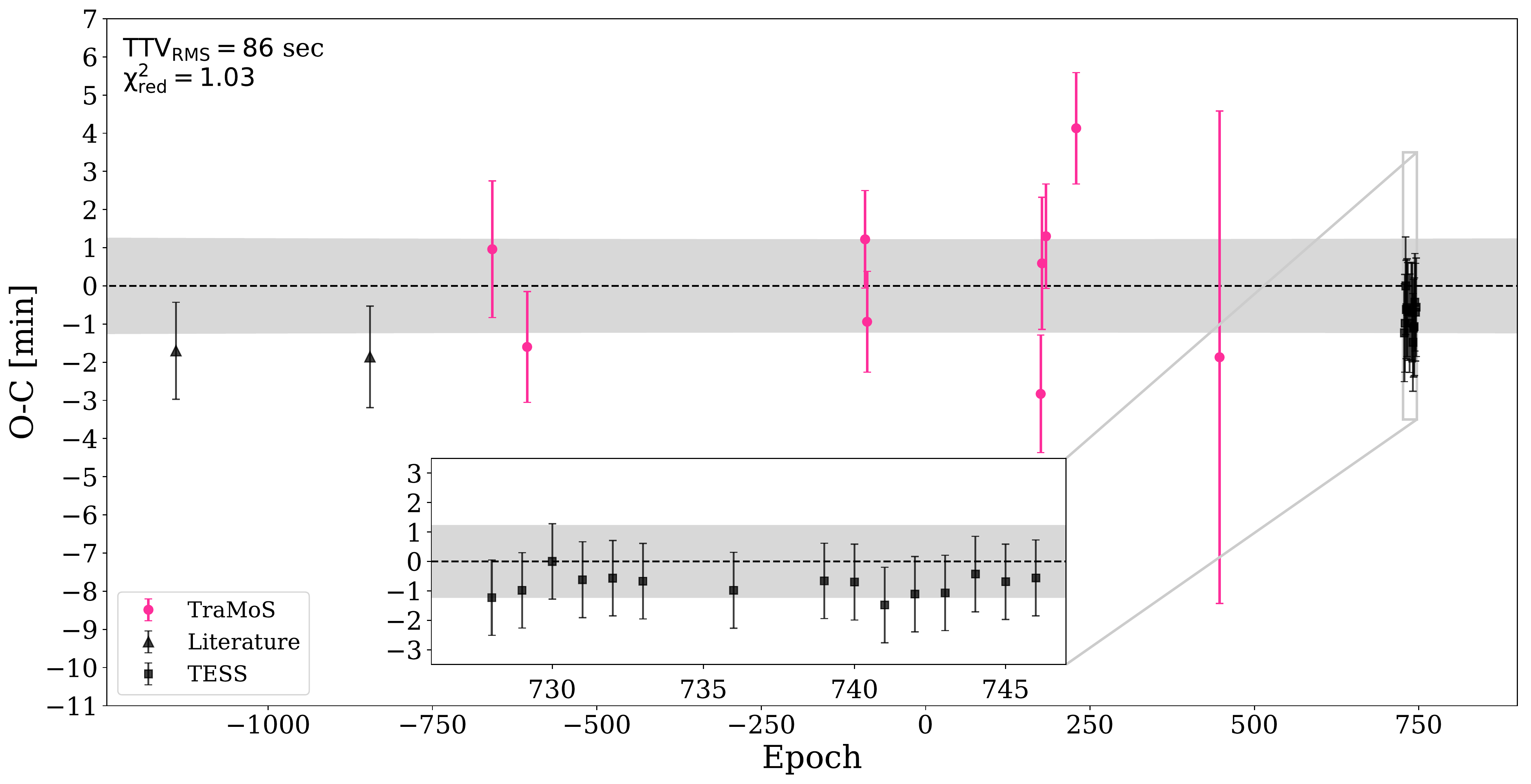}
        \caption{Observed minus calculated transit mid-times (TTV) for WASP-18Ab (\emph{top panel}), WASP-19b (\emph{center panel}) and WASP-77Ab (\emph{bottom panel}). The dashed black line corresponds to the proposed linear ephemeris, i.e. zero deviation from the predicted transit mid-time (See Section \ref{ttvsection}) computed from our refined orbital period. For that, we considered 63, 88 and 26 transit times of WASP-18Ab, WASP-19b, and WASP-77Ab, respectively. The grey area corresponds to the error propagation at $1\sigma$, where the quadratic trend looks almost horizontal. The circles in color are the TTVs from the new light curves of the TraMoS project (WASP-18Ab: light blue, WASP-19b: orange, WASP-77Ab: pink). In black are TTVs measured from different sources of transit mid-time data: the triangles are previously published transit mid-times and the squares are TESS data. The RMS scatter from the linear ephemeris are 47 seconds for WASP-18Ab; 65 seconds for WASP-19b, and 86 seconds for WASP-77Ab.}
        \label{ttv}
\end{figure*}

A refined orbital period was linearly fitted, considering a total of 63, 88 and 26 transit times of WASP-18Ab, WASP-19b, and WASP-77Ab, respectively. Along with the linear model, we also tested a second-degree polynomial to analyze a possible orbital decay. Both models considered the errors of the data. In Figure~\ref{ttv} are presented all the TTV measurements for the transit mid-times of WASP-18Ab, WASP-19b and, WASP-77Ab.

If the planet stays in a Keplerian orbit, its transit mid-time $T_{c}$ of each epoch $E$ should follow a linear function of the orbital period $P$.

\begin{equation} \label{eq1}
T_{\rm C}(E)=T_{\rm C}(0)+E \cdot P
\end{equation}

Where $T_{c}(0)$ is the optimal transit time in an arbitrary zero epoch. The best-fitted values of $T_{c}(0)$ for WASP-18Ab, WASP-19b and WASP-77Ab, are listed in Tables~\ref{wasp18},~\ref{tab:wasp19} and \ref{tab:wasp77}, respectively.

\subsubsection{WASP-18Ab}
For this system, the proposed linear ephemeris equation considering 63 transit mid-times is:

\begin{equation} \label{eq1_w18}
T_{\rm C}(E)=2456926.27460\pm(94)+E \cdot 0.941452417\pm(27)
\end{equation}

Table~\ref{times_wasp18} lists the transit mid-times and their deviation from the proposed linear ephemeris (TTV) of TraMoS data, previous published works \citep{Triaud2010,Hellier2009,Maxted2013b} and TESS light curves of WASP-18Ab.

The top panel of Figure~\ref{ttv} is the linear plot of TTV versus epoch for this planet. The deviations of the transit mid-times from the linear ephemeris has an RMS of 47 seconds. The greater deviations come from the transit mid-time of the epochs $-1904$ and $-1184$, which are over the linear ephemeris by $2.1\sigma$. If those values are removed, the RMS decreases to 35 seconds. The TTVs list in Table Table~\ref{times_wasp18}, except the epochs $-1904$ and $-1184$, lie within $1.5\sigma$ from the linear fit.

The epoch $-1184$ has the greatest error in our sample because it is not a complete transit, while the epoch $-1904$, the one with the highest deviation from the linear ephemeris on our sample, was observed during not optimal weather conditions.

When testing the goodness of the linear fit, $\chi^{2}_{red} =0.36$, while for a second-degree polynomial is $\chi^{2}_{red}=0.35$, therefore an orbital decay can be discarded in agreement with theoretical estimations \citep{CollierCameron2018}.

\subsubsection{WASP-19b} 
The proposed equation for linear ephemeris, considering 88 transit times of WASP-19b is:
\begin{equation} \label{eq1_w19}
T_{\rm C}(E)=2456402.7128\pm(16)+E \cdot 0.788838940\pm(30)
\end{equation}

The TTV values from the proposed linear ephemeris are listed in Table \ref{times_wasp19}, including transit mid-times from TraMos, previous works \citep{Hebb2010,Anderson2010,Lendl2013,Tregloan2013,Bean2013,Mancini2013} and TESS. Some of the transit mid-times from \cite{Mancini2013} come from the Exoplanet Transit Database catalog and they are accordingly identified in Table~\ref{times_wasp19}.

At the middle panel of Figure~\ref{ttv} are the TTV values versus epoch, for all the transit time considered in this work.

The RMS from the linear ephemeris is about 65 seconds. The epochs $-1311$, $-1011$, $-886$ and $-642$ have a deviation above $3\sigma$ from the linear ephemeris. If they are removed, then the RMS decreases to 52 seconds. Moreover, in our data the epoch $-611$ has one of the greatest errors due to bad weather conditions.

Considering all the transit mid-times from Table~\ref{times_wasp19}, the linear fit has $\chi^{2}_{red} = 0.65$. A second-degree polynomial was also tested to reject or not a possible orbital decay. The goodness of that fit is $\chi^{2}_{red}=0.64$.

\subsubsection{WASP-77Ab}
As in the previous targets, we computed a refined linear ephemeris equation for WASP-77A considering 26 transit times:

\begin{equation} \label{eq1_w77}
T_{\rm C}(E)=2457420.88439\pm(85)+E \cdot 1.36002866\pm(17)
\end{equation}

In Table~\ref{tab:wasp77} are listed the TTV values of our transit times (TraMoS), previous works \citep{Turner2016,Maxted2013} and TESS. At the bottom of Figure~\ref{ttv}, the TTV of WASP-77Ab is plotted versus epoch. The scatter of all the transit times is about $RMS=86$ seconds.

The epochs $175$ and $229$ are $2.3\sigma$ and $3\sigma$, respectively, above from the expected transit time following the linear ephemeris. The rest of the epochs lie within $1.5\sigma$ from it. When removing the epochs $175$ and $229$, the RMS decreases to 66 seconds.

Considering all the transit times, the linear fit has $\chi^{2}_{red}=1.03$, and the second-degree polynomial has $\chi^{2}_{red}=0.72$. Anyway, the second-order fit is highly dominated by the outlier at epoch 229 (see Figure \ref{ttv}). After removing it, the reduced chi-squared is $\chi^{2}_{red}=0.37$. In all cases, the best fit corresponds to the linear ephemeris. The variation in the transit mid-time at epoch 229 may be caused due to not optimal weather conditions ($100\%$ humidity) during the observation and the lack of good coverage in the after-transit baseline.

\subsection{Upper Mass Limits of a Hypothetical Perturber \label{addper}}

The results from our mid-transit time study (see Section~\ref{ttvsection}) allow us to infer an upper mass limit for an additional planet in each system. A perturbing planet will introduce a change in the mid-transit times of a known planet, which can be quantified by the RMS scatter around the nominal (unperturbed) linear ephemeris. The TTV effect is amplified for orbital configurations involving mean-motion resonances \citep{Agol2005, Holman2005, 2008ApJ...688..636N}. In principle, this amplification would allow the detection of a low-mass planetary perturbing body. A larger perturbation implies a larger RMS scatter around the nominal ephemeris.

The applied method follows the technique described in \citet{TEMP1,TEMP2,TEMP4}. The calculation of an upper mass limit is performed numerically via direct orbit integrations. For this task, we have modified the 
{\sc FORTRAN}-based {\sc MICROFARM}\footnote{\url{https://bitbucket.org/chdianthus/microfarm/src}}
package \citep{go2003,go2008} which utilizes {\sc OpenMPI}\footnote{\url{https://www.open-mpi.org}} to spawn hundreds of single-task parallel jobs on a suitable super-computing facility. The package's main purpose is the numerical computation of the Mean Exponential Growth factor of Nearby Orbits
\citep[MEGNO]{cincsimo2000, go2001, ci2003} over a grid of initial values of orbital parameters for an $n$-body problem. The calculation of the RMS scatter of TTVs in the present work follows a direct brute-force method, which proved to be robust given the availability of computing power.

Within the framework of the three-body problem, we integrated the orbits of one of our three hot Jupiters and an additional perturbing planet around their host stars. The mid-transit time was calculated iteratively to a high precision from a series of back-and-forth integrations once a transit of the transiting planet was
detected. The best-fit radii of both the planet and the host star were accounted for. We then calculated an
analytic least-squares regression to the time-series of transit numbers and mid-transit times to determine a
best-fitting linear ephemeris with an associated RMS statistic for the TTVs. The RMS statistic was based on a 20-year integration corresponding to 7763 transits for WASP-18b, 9270 transits for WASP-19b, and 5371 transit events for WASP-77Ab. This procedure was then applied to a grid of masses and semi-major axes of the perturbing planet while fixing all the other orbital parameters. In this study, we have chosen to start the perturbing planet on a circular orbit that is co-planar with the transiting planet; this implies that $\Omega_2=0^{\circ}$ and $\omega_2=0^{\circ}$ for the perturbing and $\Omega_1 = 0^{\circ}$ for the transiting planet. This setting provides a most conservative estimate of the upper mass limit of a possible perturber \citep{bean2009, fukui2011, Hoyer2011, Hoyer2012}. For the interested readers, we refer to \citet{TEMP4}, which has studied the effects of TTVs on varying initial orbital parameters.

\begin{table}
\caption{Approximate upper mass limits of a putative perturber in various orbital resonances for each system}
\label{masstable}
\centering
\begin{tabular}{cccc}
\hline \hline
MMR             & WASP-18A               & WASP-19                & WASP-77A              \\
($P_{2}/P_{1}$) & $[M_{\oplus}]$         & $[M_{\oplus}]$         & $[M_{\oplus}]$        \\
\hline
1:4             & -                      & -                      & 4.0                   \\
1:3             & 2.5                    & -                      & 70.0\tablefootmark{b} \\
2:5             & 1.0                    & -                      & -                     \\
1:2             & 9.0\tablefootmark{a}   & 0.26                   & 1.8                   \\
4:7             & -                      & -                      & 1.5                   \\
3:5             & -                      & -                      & 8.0                   \\
2:3             & -                      & -                      & 5.5\tablefootmark{a}  \\
11:7            & -                      & -                      & 3.0                   \\
5:3             & -                      & 2.8                    & 6.0                   \\
7:4             & -                      & -                      & 5.5                   \\
2:1             & 11.0\tablefootmark{a}  & 0.65                   & 3.0                   \\
7:3             & 6.5                    & -                      & -                     \\
5:2             & 7.5                    & 3.0                    & 105.0\tablefootmark{c}\\
3:1             & 4.0                    & 1.0                    & 50.0\tablefootmark{f} \\
17:5            & 350.0\tablefootmark{d} & -                      & -                     \\
4:1             & 7.5                    & 95.0\tablefootmark{e}  & 35.0\tablefootmark{g} \\
\hline
\end{tabular}
\tablefoot{
\tablefoottext{a}{Very close to the general instability area.}
\tablefoottext{b}{Upper mass limit from RV: 13.4 $M_{\oplus}$}
\tablefoottext{c}{Upper mass limit from RV: 26.4 $M_{\oplus}$}
\tablefoottext{d}{Upper mass limit from RV: 82.8 $M_{\oplus}$}
\tablefoottext{e}{Upper mass limit from RV: 40.8 $M_{\oplus}$}
\tablefoottext{f}{Upper mass limit from RV: 28.0 $M_{\oplus}$}
\tablefoottext{g}{Upper mass limit from RV: 30.8 $M_{\oplus}$}
}
\end{table} 

In order to calculate the location of mean-motion resonances, we have used the same code to calculate MEGNO on the same parameter grid. However, this time we integrated each initial grid point for 1000 years, allowing this study to highlight the location of weak chaotic high-order mean-motion resonances. In short, MEGNO quantitatively measures the degree of stochastic behaviour of a non-linear dynamical system and has been proven useful in the detection of chaotic resonances \citep{go2001, hi2010}. In addition to the Newtonian equations of motion, the associated variational equations of motion are solved simultaneously allowing the calculation of MEGNO at each integration time step. The {\sc MICROFARM} package implements the {\sc ODEX}\footnote{\url{https://www.unige.ch/~hairer/prog/nonstiff/odex.f}} extrapolation algorithm to numerically solve the system of first-order differential equations.

Following \citep{cincsimo2000, ci2003} the MEGNO index is defined as:
\begin{equation}
Y(t) = \frac{2}{T}\int_{0}^{T} \frac{||\dot{\boldsymbol{\delta}}(t)||}{||\boldsymbol{\delta}(t)||} t dt,
\label{eq:megno}
\end{equation}
\noindent
where $\boldsymbol{\dot{\delta}}/\boldsymbol{\delta}$ is the relative change of the variational vector $\boldsymbol{\delta}$. The time-averaged or mean of $Y(t)$ (time-averaged MEGNO) is given as:
\begin{equation}
\langle Y(t) \rangle = \frac{1}{T}\int_{0}^{T} Y(t) dt.
\label{eq:megno_mean}
\end{equation}
\noindent

The notation can be confusing at times. In \citet{cincsimo2000} the MEGNO ($Y(t)$ and $\langle Y(t) 
\rangle$ as written above) is introduced as $\mathcal{J}$ and $\bar{\mathcal{J}}$, respectively. In
\citet{ci2003} the MEGNO index and its time-average is denoted as $Y$ and $\bar{Y}$. When presenting results
(Figures~\ref{megno_wasp18} - \ref{megno_wasp77}) it is always the time-averaged MEGNO index that is utilized
to quantitatively differentiate between quasi-periodic and chaotic dynamics. The variational vector
$\boldsymbol{\delta}$ is determined from an initial-value problem by numerically solving the variational
equations of motion \cite{mikinn1999} in parallel with the Newtonian equations of motion. We refer to
\cite{hi2010} for a short and compact review of essential properties of MEGNO.

In a dynamical system that evolves quasi-periodically in time the quantity $\langle Y \rangle$ will
asymptotically approach 2.0 for $t \rightarrow \infty$. In that case, often the orbital elements associated
with that orbit are bounded. In case of a chaotic time evolution the quantity $\langle Y\rangle$ diverges away
from 2.0. with orbital parameters exhibiting erratic temporal excursions. For quasi-periodic orbits, we
typically have $|\langle Y \rangle - 2.0| < 0.001$ at the end of each integration.

Importantly, MEGNO is unable to prove that a dynamical system is evolving quasi-periodically, meaning that a
given system cannot be proven to be stable or bounded for all times. The integration of the equations of
motion only considers a limited time period. However, once a given initial condition has found to be chaotic,
there is no doubt about its erratic nature in the future.

In the following, we will present the results of each system for which we have calculated the RMS scatter of
TTVs on a grid of the masses and semi-major axes of a perturbing planet in a circular, co-planar orbit.
Results are shown in Figures~\ref{megno_wasp18} - \ref{megno_wasp77} and Table \ref{masstable}. In each of the
three cases, we find the usual instability region located in the proximity of the transiting planet with 
MEGNO color-coded as yellow (corresponding to $\langle Y\rangle > 5$). The extent of these regions coincides
with the results presented in \citet{barnes2006}. The locations of mean-motion resonances are indicated by
arrows in each map.

\subsubsection{WASP-18Ab}
For the WASP-18Ab system we find a large region of instability when
compared to the other two systems with the boundaries at the 1:2 interior
and 2:1 exterior mean-motion resonance. By over-plotting the RMS scatter
of mid-transit times ($\rm TTV_{\rm RMS}$) for a certain value, we find
that the TTVs are relatively more sensitive at orbital architectures
involving mean-motion resonances confirming the results by
\citet{Agol2005} and \citet{Holman2005}. This also applies to WASP-19 and
WASP-77A.

\begin{figure}
\includegraphics[width=1.0\columnwidth]{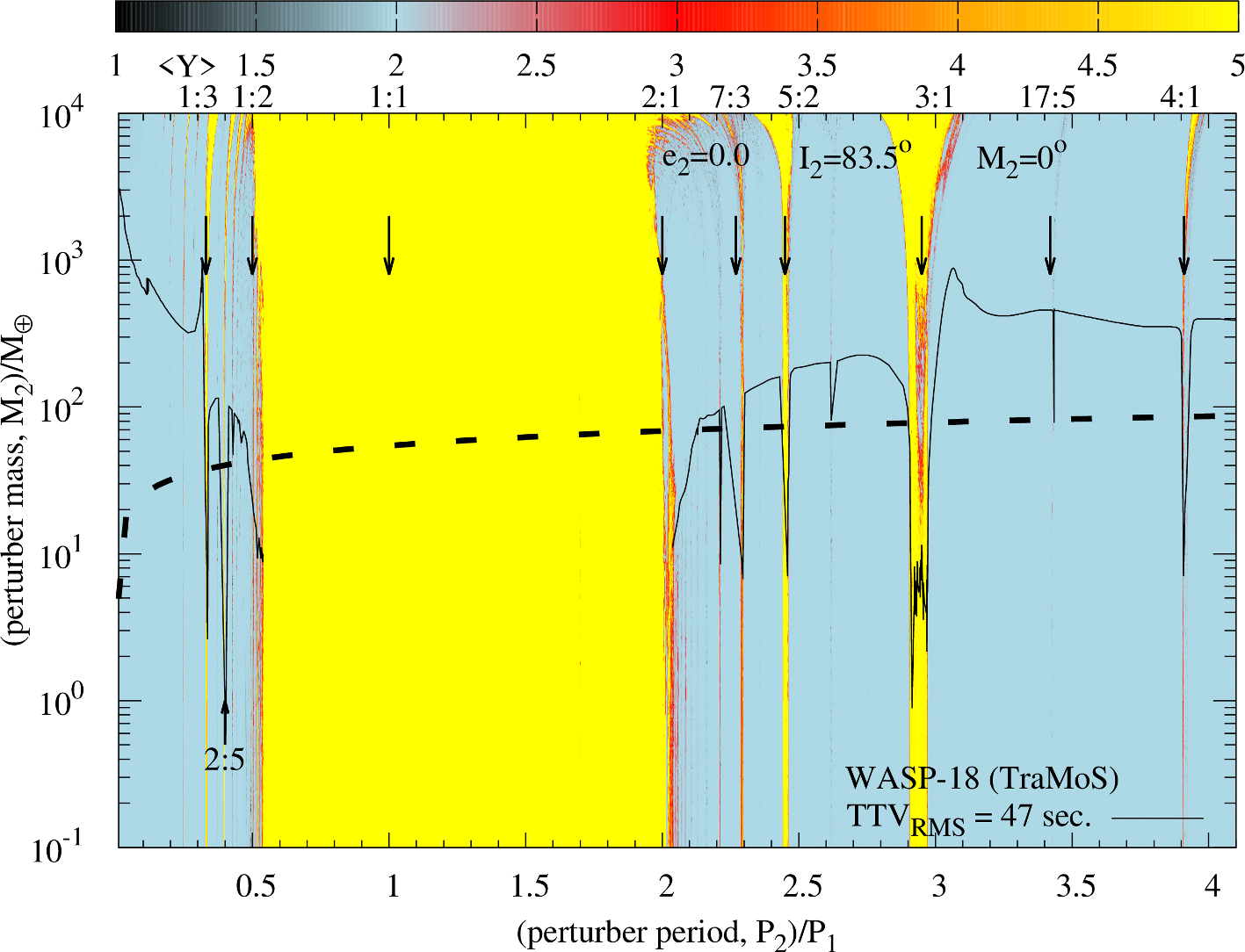}
\caption{MEGNO ($\langle Y \rangle$) stability map for the WASP-18 system. We over-plot the map with an upper mass of a hypothetical perturbing planet introducing a mid-transit time $\rm TTV_{\rm RMS}$ scatter of $47\,{\rm s}$ (solid line) as obtained in this study. The stipulated line is the upper mass limit as obtained from the RMS scatter $(30\, \rm m/s)$ of the radial-velocity curve. For initial conditions resulting in aquasi-periodic (i.e bounded) motion of the system, the $\langle Y\rangle$ value is close to 2.0 (color coded blue). For chaotic (i.e unstable) motion, the $\langle Y \rangle$ is diverging away from 2.0 (color coded red to yellow). Vertical arrows indicate $(P_2/P_1)$ orbital resonances between the perturbing body and the transiting planet. The two planets were assumed to be co-planar, and the perturbing planet's eccentricity was initially set to zero.
\emph{See electronic version for colors}.}
\label{megno_wasp18}
\end{figure}

\begin{figure}
\includegraphics[width=1.0\columnwidth]{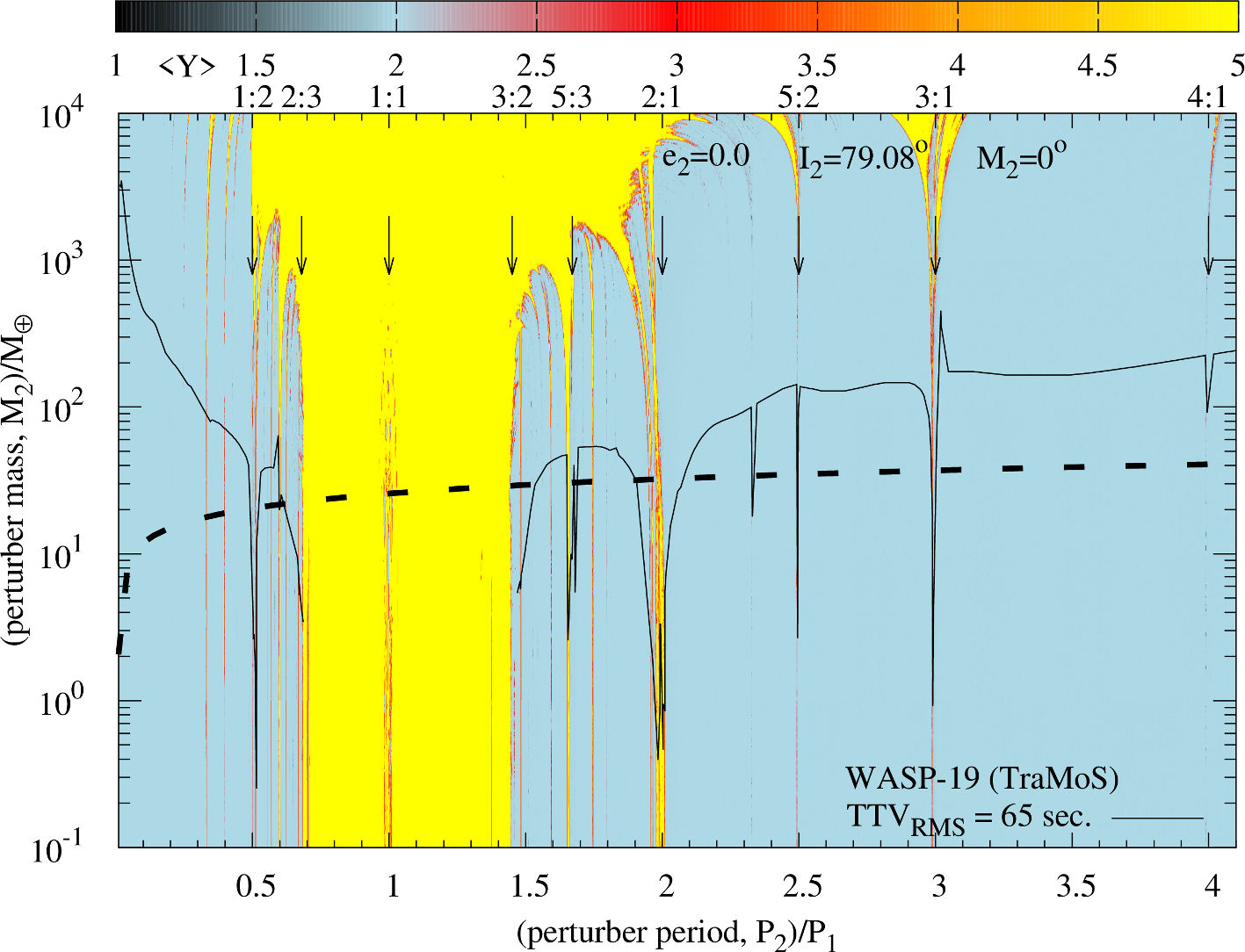}
\caption{Same as Fig.~\ref{megno_wasp18}, but this time for WASP-19 with an $\rm TTV_{\rm RMS}$ of 65 s. The
RMS for the radial-velocity measurements was $(18.2\,\rm m/s)$. \emph{See electronic version for colors}.}
\label{megno_wasp19}
\end{figure}

\begin{figure}
\includegraphics[width=1.0\columnwidth]{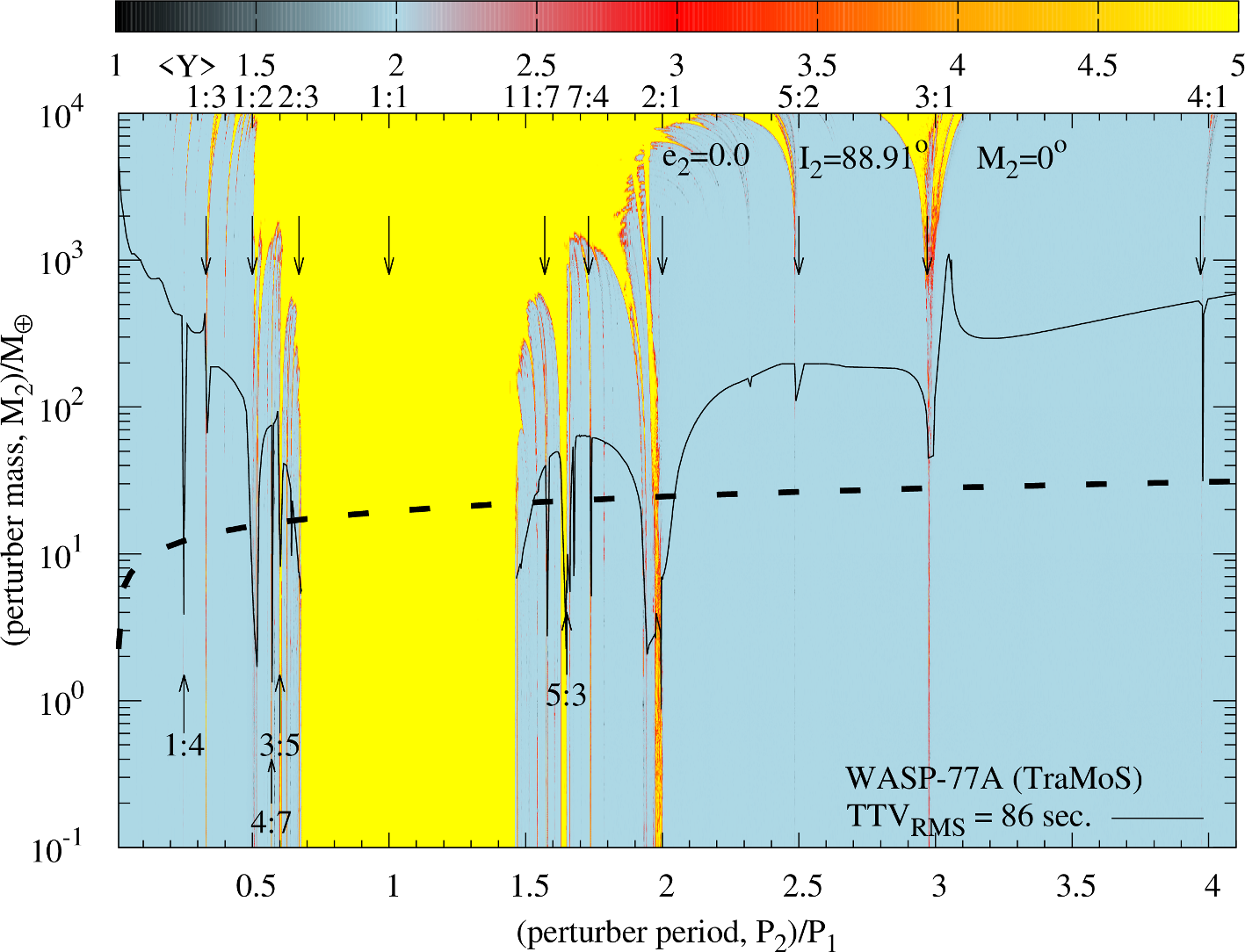}
\caption{Same as Fig.~\ref{megno_wasp18}, but this time for WASP-77 with a $\rm TTV_{\rm RMS}$ of 86 s. The
RMS for the radial-velocity measurements was $(12.0\,\rm m/s)$.
\emph{See electronic version for colors}.}
\label{megno_wasp77}
\end{figure}

As shown in Figure~\ref{megno_wasp18}, we find that a perturbing body of mass (upper limit) around $4-350~M_{\oplus}$ will cause
an RMS of $47\,{\rm s}$ when located in the $P_2/P_1=$ 7:3, 5:2, 3:1, 17:5 and 4:1 exterior resonance. For the 1:3 interior
resonance, a perturber mass (upper limit) as small as $2.5~M_{\oplus}$ could also cause a RMS mid-transit time scatter of
$47\,{\rm s}$.

Recently, \cite{Pearson2019} provided evidence for an additional perturber in the WASP-18A system with an orbital period of 2.155 days and an eccentricity of $0.009 \pm 0.006$. The mass was found to be around $50~M_{\oplus}$. When comparing this with our results, the 2.155 day period translates to a period ratio of $P_2/P_1 = 2.29$. From our dynamical analysis (see Fig. \ref{megno_wasp18}), this period ratio suggests an upper mass limit of $10~M_{\oplus}$ for a circular orbit and implies non-consistent results. At this point, we can not offer a plausible explanation for the mass difference of a factor of 5. The suggested perturber in \cite{Pearson2019} is on a near-circular orbit which coincides with our circular case. However, the difference found could be probably related to the different data set considered. While we included ground-based and TESS photometry, \cite{Pearson2019} only analysed TESS data.

\subsubsection{WASP-19b}
For the WASP-19b system the measured transit-timing RMS scatter was $\rm TTV_{\rm RMS}=65,\rm s$. Additional bodies with an upper
mass limit as low as $0.26~M_{\oplus}$ at the 1:2 (interior) mean-motion resonances could cause the observed RMS scatter.
Hypothetical planets of $2.8~M_{\oplus}$, $3.0~M_{\oplus}$ and $1.0~M_{\oplus}$ could cause the observed RMS scatter at the 5:3,
5:2 and 3:1 exterior mean-motion resonances, respectively. We refer to Fig.~\ref{megno_wasp19}.

\subsubsection{WASP-77Ab}
For the WASP-77Ab system we refer to Fig.~\ref{megno_wasp77}. The measured RMS of mid-transit timing variations around the linear
ephemeris was $\rm TTV_{\rm RMS}=86\rm s$. For interior mean-motion resonances the 1:2 and 2:3 commensurabilities could cause 
the observed $\rm TTV_{\rm RMS}$ by an additional planet of mass around $1.8~M_{\oplus}$ and $5.5~M_{\oplus}$. However, the 2:3
resonance is very close to the general instability area rendering the orbit likely to be unstable. Further a $70~M_{\oplus}$ mass
planet at the 1:3 interior resonance could also cause a $\rm TTV_{\rm RMS}$ of 86 s. A $8~M_{\oplus}$ mass planet located at the
3:5 resonance, although relatively close to the inner edge of the general instability region, could also explain the observed
timing variation. For exterior mean-motion resonances of 2:1, 3:1 and 4:1 an additional planet of mass $3.0~M_{\oplus},
50.0~M_{\oplus}$ and $35~M_{\oplus}$, could cause a $\rm TTV_{\rm RMS}=86\rm s$, respectively.

\subsection{TTV period search}

We have carried out a Lomb-Scargle period analysis \citep{Lomb1976, Scargle1982} for each system's TTVs residuals to search for a
significant periodic trend. For this we applied the
\texttt{LombScargle}\footnote{\url{http://docs.astropy.org/en/stable/stats/lombscargle.html}} (LS) algorithm available within the
\texttt{Astropy} (v3.1.1) Python package \citep{2012cidu.conf...47V, 2015ApJ...812...18V}.

The algorithm is suitable for unevenly-sampled data. We chose to carry out computations using the observed transit epochs for
each system as the independent variable. Each epoch is determined with a high degree of confidence. TTV measurement uncertainties
were not accounted for since no convincing periodic trend were detected. Default settings were avoided in order to safeguard the
analysis from an inappropriate frequency grid choice. We made use of the minimum and maximum frequency heuristic. Periods between
1 and 5000 epochs were searched for. Furthermore, we sampled each peak twelve times. Noteworthy to mention, and often overlooked,
is the possible detection of frequencies much larger than the Nyquist sampling frequency \citep{2018ApJS..236...16V}. 

\begin{figure*}
\includegraphics[width=1.0\columnwidth]{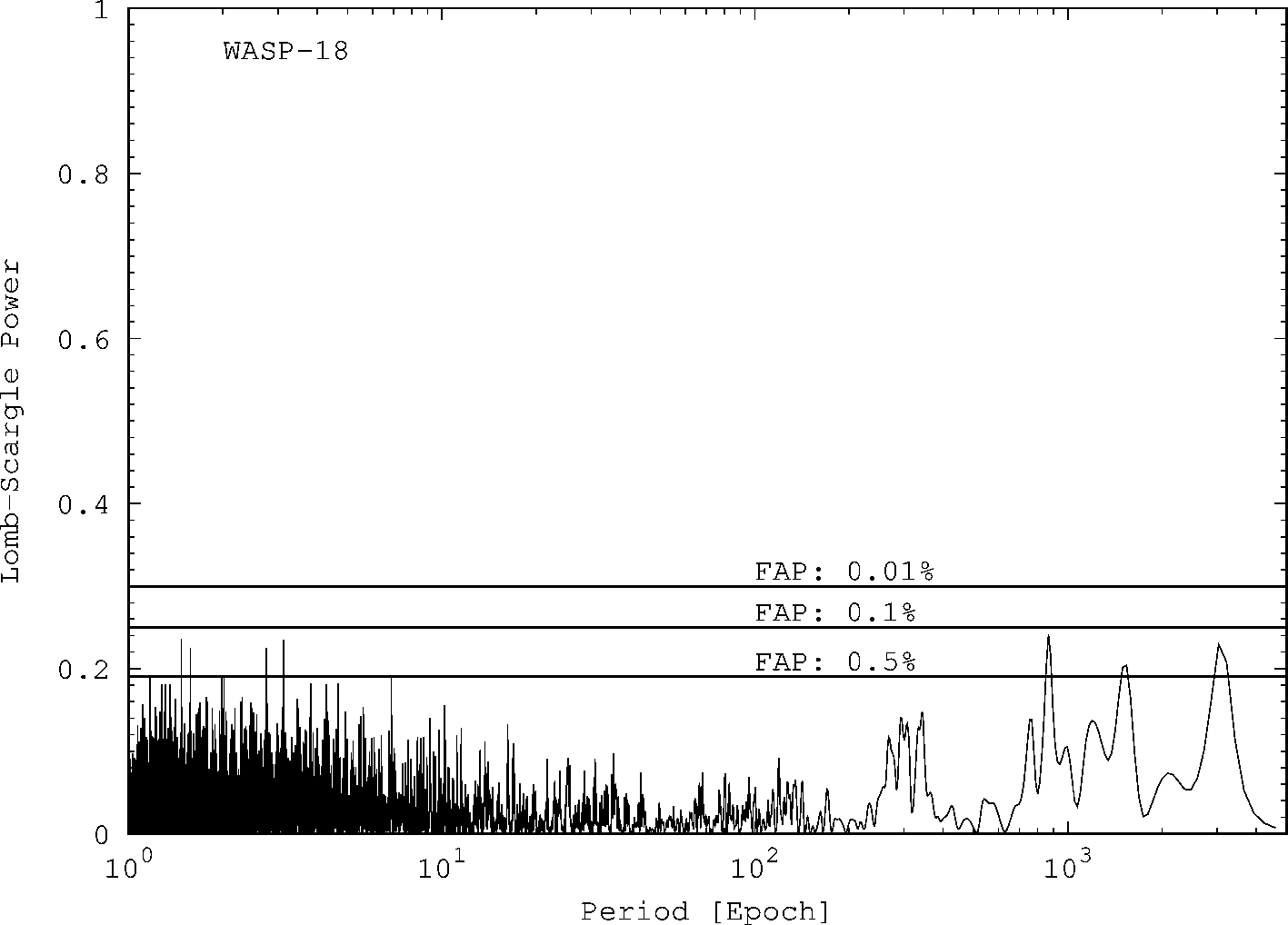}
\includegraphics[width=1.0\columnwidth]{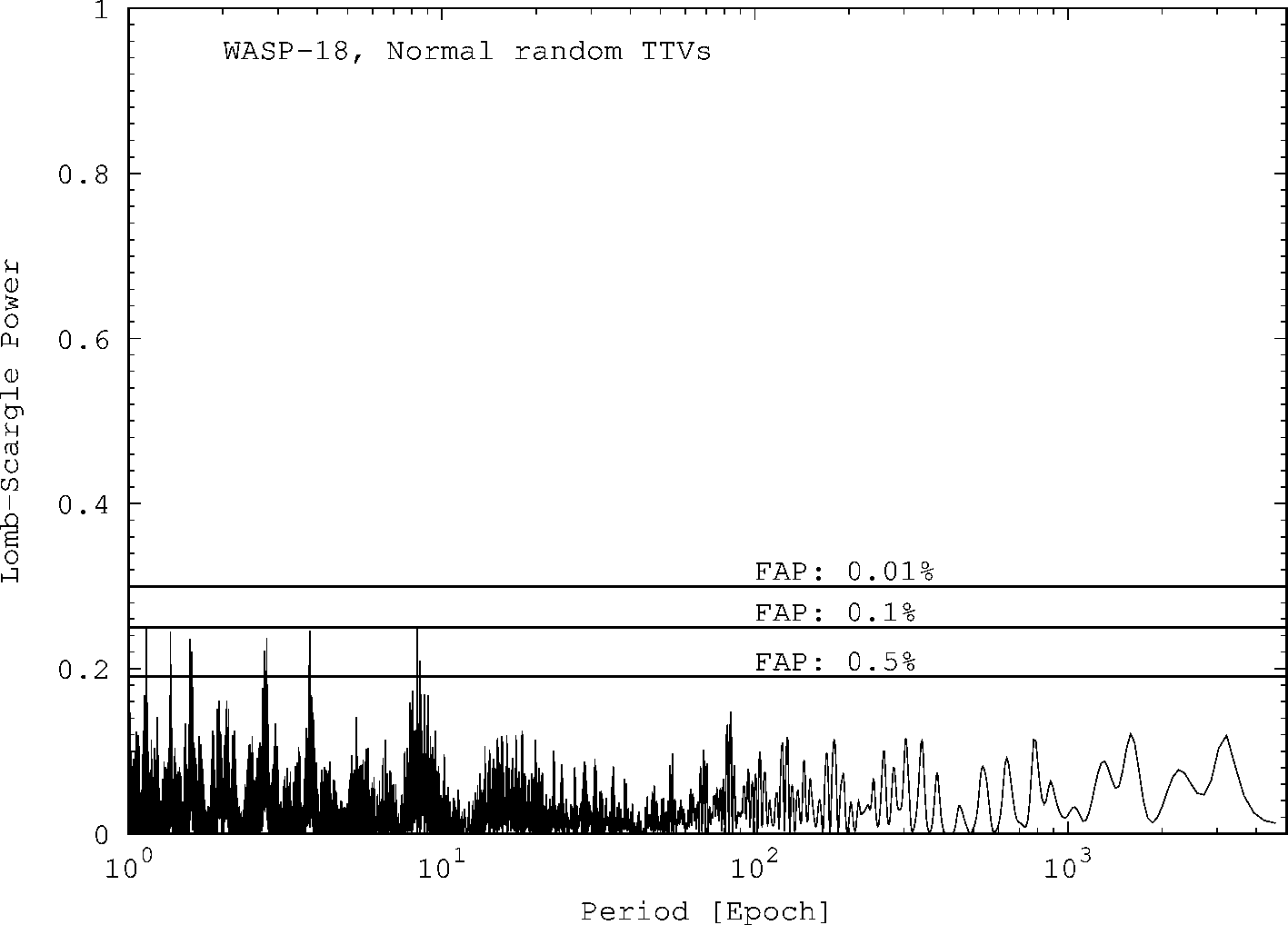}
\caption{Lomb-Scargle (standard normalized) power vs period for observed TTV residuals of WASP-18A (\emph{left
panel}) and for a simulated set of TTVs randomly drawn from a normal distribution with mean zero and standard
deviation of 0.78 minutes (\emph{right panel}). See text for more details.}
\label{LS_wasp18_random}
\end{figure*}

\begin{figure*}
\includegraphics[width=1.0\columnwidth]{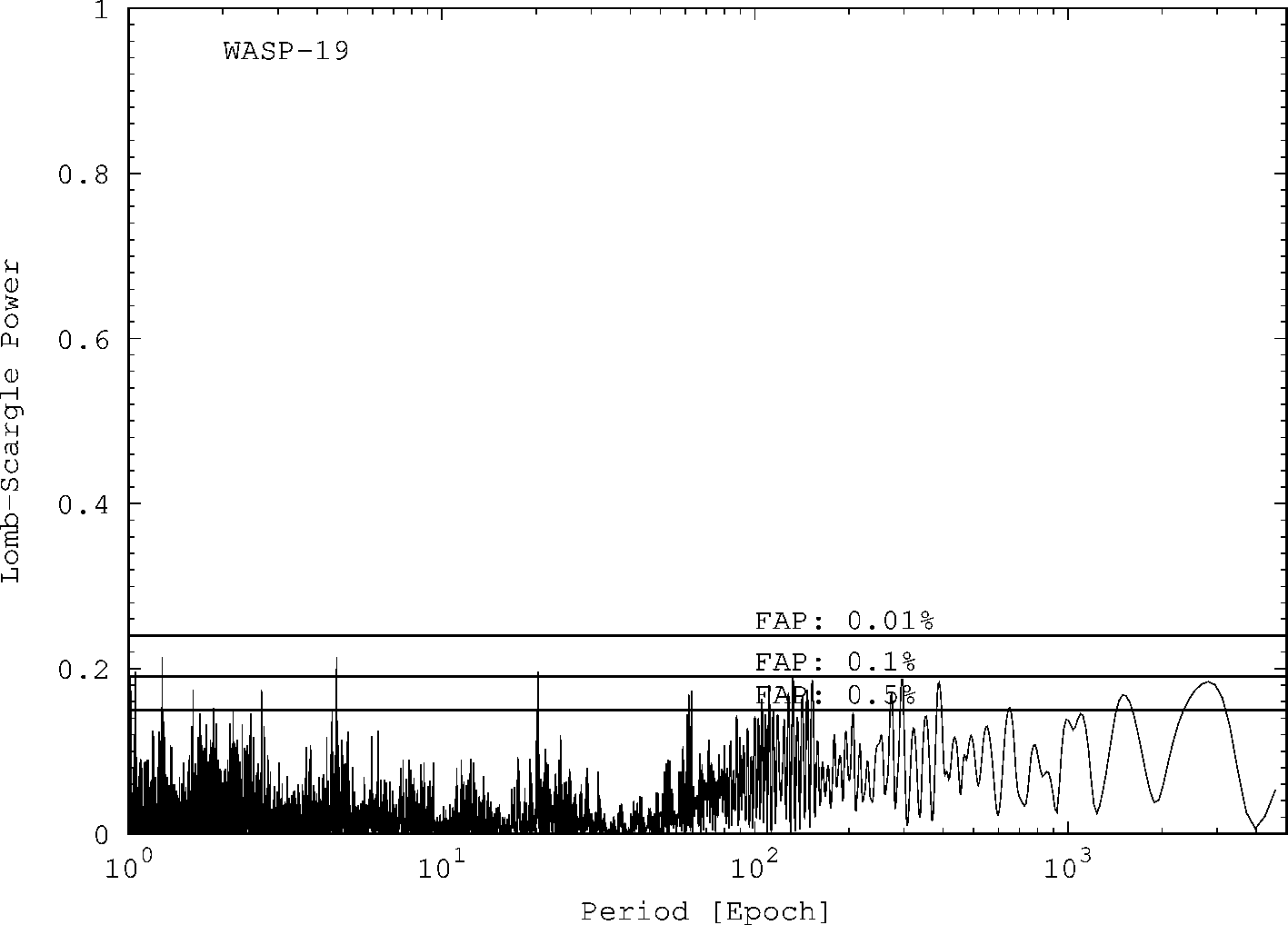}
\includegraphics[width=1.0\columnwidth]{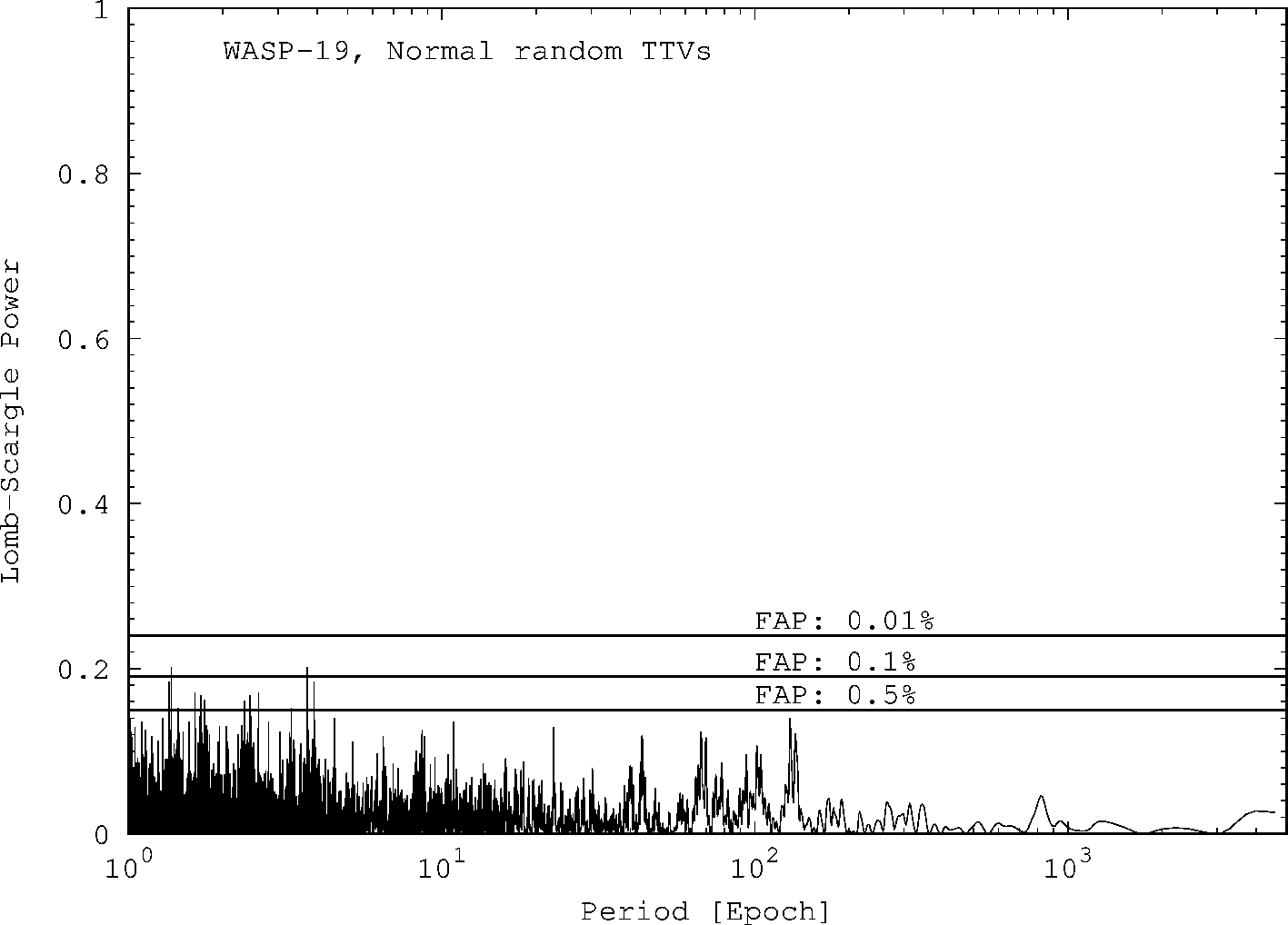}
\caption{Lomb-Scargle (standard normalized) power vs period for observed TTV residuals of WASP-19 (\emph{left
panel}) and for a simulated set of TTVs randomly drawn from a normal distribution with mean zero and standard
deviation of 1.08 minutes (\emph{right panel}). See text for more details.}
\label{LS_wasp19_random}
\end{figure*}

\begin{figure*}
\includegraphics[width=1.0\columnwidth]{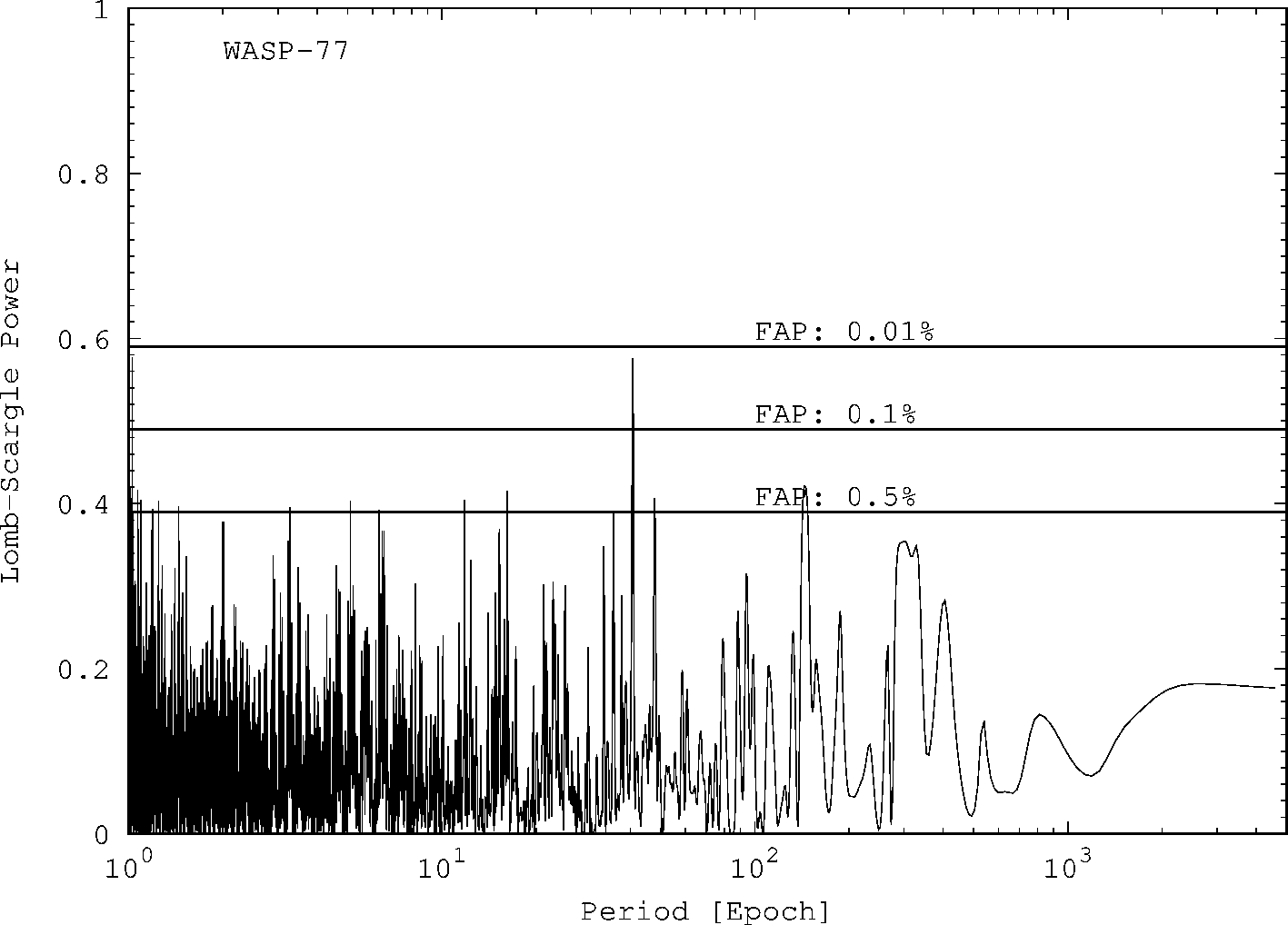}
\includegraphics[width=1.0\columnwidth]{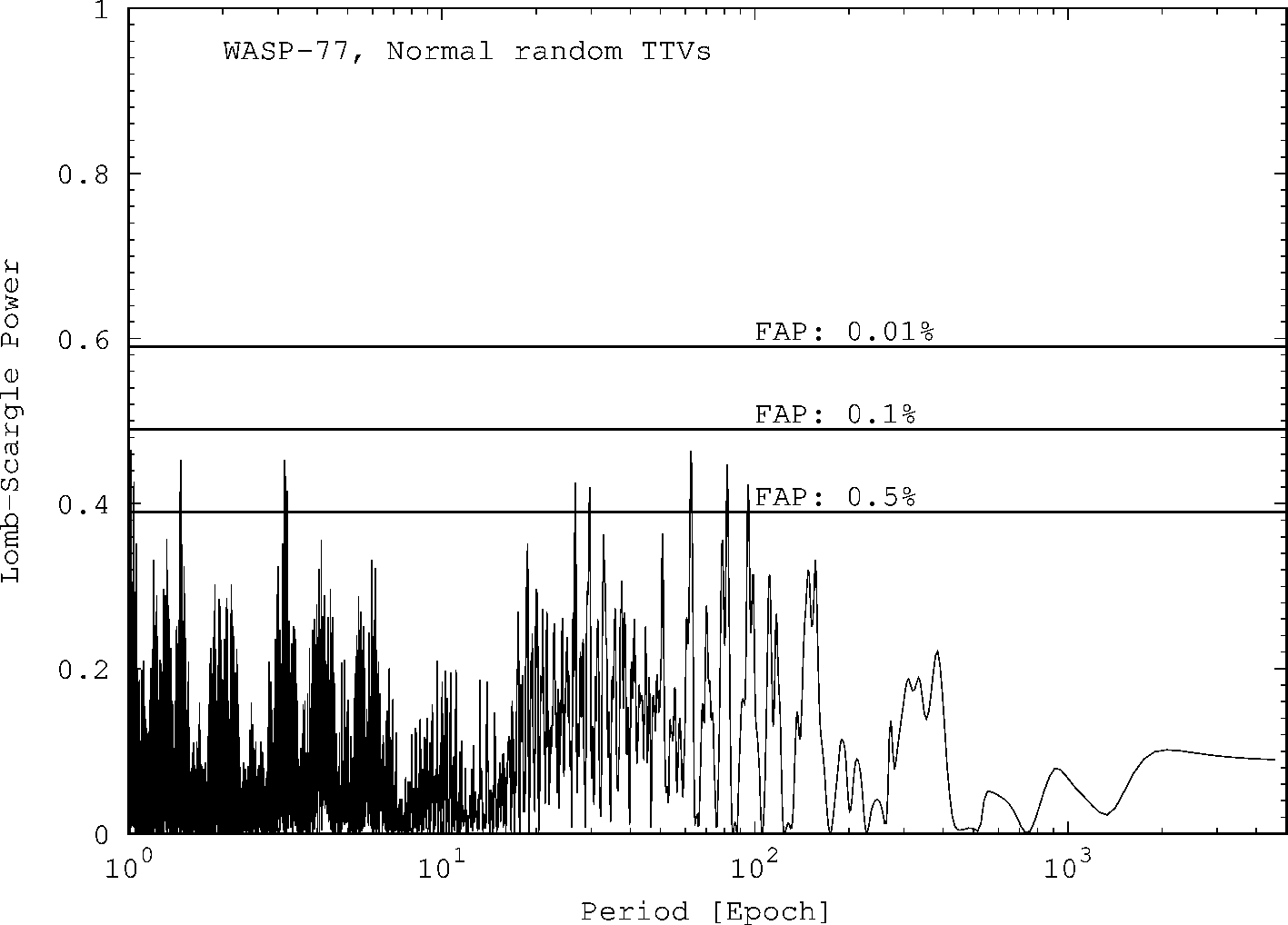}
\caption{Lomb-Scargle (standard normalized) power vs period for observed TTV residuals of WASP-77A 
(\emph{left panel}) and for a simulated set of TTVs randomly drawn from a normal distribution with mean zero
and standard deviation of 1.43 minutes (\emph{right panel}). See text for more details.}
\label{LS_wasp77_random}
\end{figure*}

The result for each system is shown in Figs.~\ref{LS_wasp18_random} to \ref{LS_wasp77_random}, were we show the Lomb-Scargle
power $P$ from the standard normalization method with $0\le P<1$. The final period is found by multiplying with the final
best-fit period for each system. To quantify the significance of period-peaks we calculated the false-alarm probability (FAP) for
three different $p$-values. The FAP encodes the probability of detecting a peak of a given height (or higher) and is conditioned
on the null-hypothesis that the data is characterized by normal random noise.

To avoid misinterpretation of the FAP we have calculated synthetic random TTVs for each system in a single realization. For each
known epoch, we drew a normal random point with mean zero and standard deviation in accordance with the measured RMS for each
timing data set (47 s for WASP-18A, 65 s for WASP-19 and 86 s for WASP-77A).

We then recomputed the LS periodogram for each synthetic data set. This method enables a meaningful quantitative assessment of a
minimum requirement of the FAP to detect a true periodic signal which clearly stands out from Gaussian noise. We plot the LS
periodograms for the synthetic TTVs in the right panels of Figs.~\ref{LS_wasp18_random} to \ref{LS_wasp77_random}. For all three
systems we find that a reasonable minimum FAP of 0.1\% is required in order to distinguish any true signal in our data from 
white noise. In generally, for all three systems, we found no significant (99.99\% level) periodicity peaks with a FAP of 
0.01\% or smaller. The only system that exhibits a period with a $\rm FAP \simeq 0.01\%$ is WASP-77A for which a period at 
40 epochs was found corresponding to $P_2 \simeq 40 \times 1.36\rm\,days \simeq 54\rm\,days$.

\section{Summary and Conclusions}\label{summary}

We performed a photometric follow-up of the transiting exoplanets WASP-18Ab, WASP-19b, and WASP-77Ab using meter-class telescopes within the TraMoS project. Our 22 new high-precision light curves and archival data were combined to refine the physical and orbital parameters of the systems.

For WASP-18Ab we find a larger value for the fraction of the radius $R_p/R_*$ than the most recent work with TESS data \cite{Shporer2018}, and a larger total transit duration $T_{14}$ comparing with \cite{Hellier2009}. The rest of the stellar and planetary parameters are all in good agreement with previous results.

In the analysis of WASP-19b, our results are in general, in good agreement with previous works \citep{Hebb2010,Lendl2013}. Only the inclination $i$ and the total duration of the transit $T_{14}$ show important differences.

In this work, we reported the first bulk measurements of the WASP-77Ab system. We find almost no disagreement in the orbital and physical parameters with the discovery paper \cite{Maxted2013}.

We included archival transit mid-times and new TESS light curves along with the transits from the TraMoS project, to obtained refined values for the period $P$ of the three exoplanets. We report an orbital period of $P=0.941452417\pm2.7\times10^{-8}$ days for WASP-18Ab, $P=0.788838940\pm3\times10^{-9}$ days for WASP-19b, and $P=1.36002866\pm1.7\times10^{-7}$ days for WASP-77Ab. With these refined orbital periods, we proposed updated linear ephemeris for the three targets. The scatter in the transit mid-time $\rm TTV_{RMS}$ is 47 seconds, 65 seconds, and 86 seconds, for WASP-18Ab, WASP-19b, and WASP-77Ab, respectively. Also, we found a lack of significant TTV periodic signals.  

The $\rm TTV_{RMS}$ could be produced by a perturber body gravitationally bounded with our targets. Thus, we performed orbit integrations to find upper mass limits for possible companions. We found that, for WASP-18Ab, the observed RMS could be produced by a perturber with an upper limit mass of around $4-7.5~M_{\oplus}$ in 7:3, 5:2, 3:1, and 4:1 exterior resonances, and for the interior resonances 1:3, 1:2, 2:1, and 2:5, a body with a mass between $1-11~M_{\oplus}$. We compared our results with the recent submission of \cite{Pearson2019}, where evidence of a possible perturber of $50~M_{\oplus}$ with an orbital period of 2.155 days is presented. However, for that period, our results place an upper mass limit of $10~M{\oplus}$. 

In the case of WASP-19b, companions with upper limit masses between $0.65-3~M_{\oplus}$ in 2:1, 5:2, 5:3, and 3:1 exterior resonances, could produce the 75 seconds of scatter. As well as a $0.26~M_{\oplus}$ body in 1:2 interior resonance.

For WASP-77Ab, the observed RMS in the TTVs could be produced by planets with masses between $1.5-8~M_{\oplus}$ in the interior resonances 1:2, 3:5, 2:3, 1:4, and 4:7. And for exterior resonances 11:7, 5:3, 7:4, and 2:1, perturbing bodies with masses between $3-6~M_{\oplus}$.

The hypothetical perturbers with the greatest masses for the three targets are discarded, as they are constrained by RV variations. These cases are: a body up to $350~M_{\oplus}$ in 17:5 resonance for WASP-18Ab, $95~M_{\oplus}$ in 4:1 resonance for WASP-19b, and $70~M_{\oplus}$, $50~M_{\oplus}$, $105~M_{\oplus}$,  and $35~M_{\oplus}$ in 1:3 3:1, 5:2, 4:1, and resonances, respectively, for WASP-77Ab. The possible perturbers presented in this work, that are not discarded by RV limits, does not exceed $11~M_{\oplus}$.

We find no significant periodicity in the TTV curves of the three exoplanets by performing a Lomb-Scargle period analysis.

At this point, we find no evidence that a second-degree model is better than a linear model for WASP-18Ab. This supports the conclusion that there is no evidence for a rapid orbital decay, as proposed by \cite{Wilkins2017}. As the TTV technique is sensitive to detect tidal decays on the exoplanets' orbits, we could detect any trending in the TTV data, which is not the case. Moreover, theoretical studies \citep{CollierCameron2018} suggest a time of around 20 years to observe a variation of 4 seconds in this system. Our results support that prediction.

Previous photometric studies of WASP-19b \citep{Lendl2013,Wong2016} also suggest the lack of TTV on this system. Our results include more transit times, 88 versus 56 in \cite{Wong2016} and 14 in \cite{Lendl2013}, and also more recent transits light curve coming from TESS. Finding a no periodic TTV signal is consistent with their results.

This is the first detailed study of WASP-77Ab. Our results will serve as a base for future photometric and dynamic studies where an extensive follow-up should be performed. WASP-77Ab shows the larger deviation for the linear ephemeris of our targets, with almost 1.5 minutes. More consecutive transit times are needed to understand the true nature of this planet and its possible companions.

The Kepler mission provided continuous photometric monitoring of thousand of stars, that ended up with the first discovery of a planetary system showing TTVs. Now TESS with its observing plan divided into sectors is delivering an important amount of photometry data especially for short-period exoplanets, as hot Jupiters. Combining the new TESS data with ground-based follow-up observations, many possible short-period TTVs could be confirmed or ruled out.

To date, none of the previous and current targets of the TraMoS project have shown TTVs \citep{Hoyer2016,Hoyer2013,Hoyer2012}. As they are all hot Jupiters, these results suggest that probably, this kind of planets are isolated on their systems or accompanied by small bodies, making difficult to detect them. How WASP-47b \citep{Becker2015} and Kepler-730b \citep{Canas2019}, the only two hot Jupiters showing TTVs, have close-in companions is still unknown. Moreover, none of the current formation theories of this kind of exoplanets, predict the occurrence ratio of close-in companions in their systems. However, \cite{Steffen2012} analyzed Kepler data to constraint the occurrence rate of companion in hot Jupiter systems. In a sample of 63 candidates, none of them show evidence of close-in companions. Thus, continuing performing photometric follow-up of hot Jupiters is crucial to unveil their planetary formation process.

\begin{acknowledgements}
    The authors thank the referee for the helpful comments that improved the quality of this publication. We also acknowledge the Exoplanet Transit Database (ETD) observers: J. Gonzalez, P. Benni, D. Molina and J. Gaitan for providing us their observations of WASP-77Ab. The authors thank Ian Wong and Avi Shporer for their insights during the preparation of this paper. TCH acknowledges Dr. Jennifer Burt for valuable discussion on the computation of Lomb-Scargle power spectra within {\sc PYTHON}. PCZ thanks the Graduate Department, Vice-Presidency of Academic Affairs of the Universidad de Chile, and Yale University for their support in the Short Term Research Programme. PCZ also thanks the LSSTC Data Science Fellowship Program, which is funded by LSSTC, NSF Cybertraining Grant \#1829740, the Brinson Foundation, and the Moore Foundation; her participation in the program has benefited this work. PR acknowledges support from CONICYT project Basal AFB-170002. SW thanks Heising-Simons Foundation for their generous support. 
\end{acknowledgements}

\bibliographystyle{aa}
\bibliography{references}

\clearpage
\onecolumn
\longtab{
\begin{longtable}{cccc}
\caption{\label{times_wasp18} Transit mid-times of WASP-18Ab}\\
\hline\hline
Epoch & Transit mid-time & TTV & References\\
      & (${\rm BJD_{TDB}}$) & (min) & \\
\hline 
\endfirsthead
\caption{Transit mid-times of WASP-18Ab continued}\\
\hline\hline
Epoch & Transit mid-time & TTV & References\\
      & (${\rm BJD_{TDB}}$) & (min) & \\
\hline
\endhead
\hline
\endfoot
-2873 & $2454221.48238$ & $0.1\pm1.5$ & \citet{Hellier2009}\\
-2402 & $2454664.9061$ & $-0.4\pm1.4$ & \citet{Triaud2010} \\
-1904 & $2455133.7472$ & $-3.4\pm1.7$ & TraMoS \\
-1903 & $2455134.6914$ & $0.6\pm1.7$ & TraMoS \\
-1902 & $2455135.6331$  & $0.9\pm1.7$  & TraMoS \\
-1811 & $2455221.3042$ & $-0.6\pm1.4$  & \citet{Maxted2013}\\
-1629 & $2455392.6474$ & $-2.2\pm1.5$& \citet{Maxted2013} \\
-1601 & $2455419.0083$ & $-1.8\pm2.2$& \citet{Maxted2013}\\
-1587 & $2455432.1897$ & $-0.3\pm1.4$ & \citet{Maxted2013}\\
-1546 & $2455470.7885$ & $-1.4\pm1.4$ & \citet{Maxted2013}\\
-1543 & $2455473.6144$ & $0.9\pm1.9$& \citet{Maxted2013}\\
-1457 & $2455554.5786$ & $-0.2\pm1.5$& \citet{Maxted2013} \\
-1440 & $2455570.5842$ & $1.2\pm1.6$& \citet{Maxted2013}\\
-1184 & $2455811.5970$ &  $2.7\pm5.9$ & TraMoS  \\
-1115 & $2455876.5559$ & $0.8\pm2.3$ & \citet{Maxted2013} \\ 
776 & $2457656.84078$ & $-1.1\pm1.4$ & TraMoS  \\
777 & $2457657.78359$   & $0.9\pm1.4$ & TraMoS\\
778 & $2457658.72404$ & $-0.6\pm1.4$ & TraMoS \\
1169 & $2458026.83186$ & $-0.6\pm1.5$ & TraMoS\\
1518 & $2458355.39936$ & $ 0.1\pm0.9$ & TESS\\
1519 & $2458356.34074$ & $ 0.0\pm0.9$ & TESS\\
1520 & $2458357.28226$ & $ 0.1\pm0.9$ & TESS\\
1521 & $2458358.22348$ & $ -0.3\pm 0.9$ & TESS\\
1522 & $2458359.16512$ & $ 0.0\pm0.9$ & TESS\\
1523 & $2458360.10663$ & $ 0.1\pm0.9$ & TESS\\
1524 & $2458361.04812$ & $ 0.1\pm 0.9$ & TESS\\
1525 & $2458361.98970$ & $ 0.3\pm0.9$ & TESS\\
1526 & $2458362.93130$ & $ 0.5\pm 0.9$ & TESS\\
1527 & $2458363.87265$ & $ 0.4\pm 0.9$ & TESS\\
1528 & $2458364.81372$ & $ -0.2\pm0.9$ & TESS\\
1529 & $2458365.75526$ & $ 0.0\pm 0.9$ & TESS\\
1530 & $2458366.69705$ & $ 0.5\pm 0.9$ & TESS\\
1533 & $2458369.52129$ & $ 0.3\pm 0.9$ & TESS\\
1534 & $2458370.46287$ & $ 0.5\pm 0.9$ & TESS\\
1535 & $2458371.40404$ & $ 0.1\pm 0.9$ & TESS\\
1536 & $2458372.34538$ & $ -0.1 \pm0.9$ & TESS\\
1537 & $2458373.28724$ & $ 0.5\pm 0.9$ & TESS\\
1538 & $2458374.22817$ & $ -0.3\pm 0.9$ & TESS\\
1539 & $2458375.16977$ & $ 0.0 \pm0.9$ & TESS\\
1540 & $2458376.11119$ & $ -0.1 \pm 0.9$ & TESS\\
1541 & $2458377.05269$ & $ 0.0\pm 0.9$ & TESS\\
1542 & $2458377.99449$ & $ 0.5\pm 0.9$ & TESS\\
1543 & $2458378.93575$ & $ 0.2\pm 0.9$ & TESS\\
1544 & $2458379.87717$ & $ 0.2\pm 0.9$ & TESS\\
1551 & $2458386.46725$ & $ 0.0\pm 0.9$ & TESS\\
1552 & $2458387.40886$ & $ 0.3\pm 0.9$ & TESS\\
1553 & $2458388.35016$ & $ 0.0\pm 0.9$ & TESS\\
1554 & $2458389.29158$ & $ 0.0\pm 0.9$ & TESS\\
1555 & $2458390.23338$ & $ 0.5\pm 0.9$ & TESS\\
1556 & $2458391.17453$ & $ 0.1\pm 0.9$ & TESS\\
1557 & $2458392.11596$ & $ 0.0\pm 0.9$ & TESS\\
1558 & $2458393.05747$ & $ 0.1\pm 0.9$ & TESS\\
1559 & $2458393.99900$ & $ 0.2\pm 0.9$ & TESS\\
1560 & $2458394.94027$ & $ 0.0\pm 0.9$ & TESS\\
1562 & $2458396.82304$ & $ -0.2\pm0.9$ & TESS\\
1563 & $2458397.76450$ & $ -0.2\pm 0.9$ & TESS\\
1564 & $2458398.70659$ & $ 0.7\pm 0.9$ & TESS\\
1565 & $2458399.64752$ & $ -0.1\pm 0.9$ & TESS\\
1566 & $2458400.58902$ & $ 0.0\pm 0.9$ & TESS\\
1567 & $2458401.53081$ & $ 0.5\pm 0.9$ & TESS\\
1568 & $2458402.47199$ & $ 0.1\pm 0.9$ & TESS\\
1569 & $2458403.41362$ & $ 0.4\pm 0.9$ & TESS\\
1570 & $2458404.35477$ & $ -0.1\pm 0.9$ & TESS\\
1571 & $2458405.29598$ & $ -0.4\pm 0.9$ & TESS\\

\end{longtable}   
}

\longtab{
\begin{longtable}{cccc}
\caption{\label{times_wasp19} Transit mid-times of WASP-19b}\\
\hline \hline
Epoch & Transit mid-time & TTV & References\\
      & (${\rm BJD_{TDB}}$) & (min) &  \\
\hline
\endfirsthead
\caption{Transit mid-times of WASP-19b continued}\\
\hline\hline
Epoch & Transit mid-time & TTV & References\\
      & (${\rm BJD_{TDB}}$) & (min) &  \\
\hline
\endhead
\hline
\endfoot
-2063 & $2454775.3372$ & $-1.6\pm3.2$ & \citet{Hebb2010}\\
-2061 & $2454776.91566$ & $-0.5\pm2.3$ & \citet{Anderson2010} \\
-2010 & $2454817.14633$ & $-0.6\pm2.3$ & \citet{Lendl2013} \\
-1525 & $2455199.73343$ & $-0.2\pm2.6$& Exoplanet Transit Database\\
-1459 & $2455251.79657$ & $-0.6\pm2.3$& \citet{Tregloan2013} \\
-1458 & $2455252.58544$ & $-0.5\pm2.3$& \citet{Tregloan2013} \\
-1454 & $2455255.74077$ & $-0.6\pm2.3$& \citet{Tregloan2013} \\
-1449 & $2455259.68448$ & $-1.3\pm2.4$& Exoplanet Transit Database \\ 
-1431 & $2455273.88282$ & $-2.4\pm2.5$& Exoplanet Transit Database \\
-1399 & $2455299.12768$ & $0.6\pm2.4$& Exoplanet Transit Database \\
-1354 & $2455334.6254$ & $0.5\pm2.3$ & \citet{Mancini2013}\\
-1349 & $2455338.56927$ & $0.1\pm2.3$& \citet{Lendl2013} \\ 
-1330 & $2455353.55659$ & $-0.8\pm2.3$& \citet{Mancini2013}\\
-1317 & $2455363.81131$ & $-1.1\pm2.4$& Exoplanet Transit Database  \\
-1311 & $2455368.54285$ & $-3.3\pm3.8$& \citet{Mancini2013} \\ 
-1094 & $2455539.72327$ & $0.2\pm2.4$ & \citet{Lendl2013} \\
-1056 & $2455569.69826$ & $-1.1\pm2.4$ & \citet{Lendl2013}\\
-1039 & $2455583.10979$ & $0.8\pm2.6$ & Exoplanet Transit Database \\
-1037 & $2455584.68693$ & $0.0\pm2.3$ & \citet{Lendl2013} \\
-1025 & $2455594.15188$ &  $-1.6\pm3.3$ & \citet{Mancini2013} \\
-1016 & $2455601.25164$ & $-1.3\pm2.5$ & \citet{Mancini2013}\\
-1014 & $2455602.83138$ & $1.7\pm2.4$ & \citet{Lendl2013} \\
-1011 & $2455605.19414$ & $-3.8\pm3.5$ & \citet{Mancini2013} \\
-1009 & $2455606.77464$ & $0.3\pm2.3$ & \citet{Lendl2013} \\
-1008 & $2455607.56241$ & $-1.2\pm2.4$ & \citet{Lendl2013} \\
-989 & $2455622.55057$ & $-0.9\pm2.3$ & \citet{Lendl2013} \\
-987 & $2455624.12787$ & $-1.5\pm3.1$ & \citet{Mancini2013} \\
-976 & $2455632.80612$ & $0.0\pm2.3$ & \citet{Lendl2013} \\
-947 & $2455655.68222$ & $-0.3\pm2.4$ & \citet{Lendl2013} \\
-928 & $2455670.66976$ & $-0.9\pm2.5$ & \citet{Lendl2013} \\
-923 & $2455674.61367$ & $-1.3\pm2.4$ & TraMoS \\
-919 & $2455677.77038$ & $0.7\pm3.6$ & \citet{Mancini2013} \\
-905 & $2455688.81201$ & $-2.4\pm5.3$ & \citet{Mancini2013} \\
-904 & $2455689.60276$ & $0.4\pm2.4$ & \citet{Mancini2013} \\
-900 & $2455692.75674$ & $-1.6\pm4.3$ & \citet{Mancini2013} \\
-899 & $2455693.54639$ & $-0.5\pm2.3$ & \citet{Mancini2013}  \\
-886 & $2455703.79933$ & $-3.3\pm6.4$ & \citet{Mancini2013} \\
-885 & $2455704.59078$ & $0.5\pm2.4$ & \citet{Mancini2013} \\
-880 & $2455708.534626$ & $0.0\pm 2.3$ & \citet{Mancini2013}  \\
-654 & $2455886.81234$ &  $0.2\pm3.8$ & \citet{Mancini2013} \\
-642 & $2455896.27611$ & $-3.1\pm3.8$ & \citet{Mancini2013} \\
-618 & $2455915.20980$ & $-0.9\pm2.5$ & \citet{Mancini2013}  \\
-613 & $2455919.15485$ & $0.4\pm 2.7$ & \citet{Mancini2013}\\
-611 & $2455920.7353$ & $0.0\pm5.4$ & TraMoS \\
-609 & $2455922.30966$ & $-0.4\pm8.3$ & \citet{Mancini2013} \\
-511 & $2455999.6163$ & $0.2\pm2.3$ & \citet{Bean2013} \\
-483 & $2456021.70374$ & $0.1\pm2.3$ & \citet{Bean2013} \\
-473 & $2456029.5925$ & $0.7\pm 2.4$ &\citet{Lendl2013} \\
-468 & $2456033.53643$ &  $0.3\pm 2.3$ & \citet{Mancini2013} \\
-430 & $2456063.51174$ & $-0.5\pm 2.3$ & \citet{Lendl2013} \\
-86 & $2456334.87208$ & $-0.8\pm2.4$ & Exoplanet Transit Database \\
-47 & $2456365.6373$  & $-0.1\pm2.3$& TraMoS \\
1 & $2456403.50158$ &  $-0.1\pm2.3$ & TraMoS \\
729 & $2456977.77722$  & $1.3\pm2.3$ & \citet{Sedaghati2015} \\
867 & $2457086.63571$ & $-0.5\pm2.3$& TraMoS \\
1383 & $2457493.67676$   & $-0.2\pm2.3$& TraMoS  \\
1771 & $2457799.74612$ &  $-0.3\pm2.3$ & TraMoS \\
1838 & $2457852.597807$ & $-0.7\pm2.3$ & TraMoS\\ 
2064 & $2458030.8751$  & $-1.5\pm3.2$& TraMoS \\
2716 & $2458545.19919$  & $ -0.3\pm1.0$ & TESS \\
2717 & $2458545.98937$  & $ 1.7\pm 1.0$ & TESS \\
2718 & $2458546.77766$  & $ 0.9\pm 1.0$ & TESS \\
2719 & $2458547.56690$  & $ 1.5\pm 1.1$ & TESS \\
2720 & $2458548.35441$  & $ -0.4\pm 1.0$ & TESS \\
2721 & $2458549.14447$  & $ 1.3\pm 1.0$ & TESS \\
2722 & $2458549.93234$  & $ -0.1\pm 1.0$ & TESS \\
2723 & $2458550.72134$  & $ 0.2\pm 1.0$ & TESS \\
2724 & $2458551.50995$  & $ -0.2\pm 1.0$ & TESS \\
2725 & $2458552.29885$  & $ -0.1\pm 1.0$ & TESS \\
2726 & $2458553.08808$  & $ 0.5\pm 1.0$ & TESS \\
2727 & $2458553.87683$  & $ 0.4\pm 0.9$ & TESS \\
2728 & $2458554.66545$  & $ 0.0\pm 1.0$ & TESS \\
2729 & $2458555.45414$  & $ -0.2\pm 1.0$ & TESS \\
2732 & $2458557.82191$  & $ 1.6\pm 1.0$ & TESS \\
2733 & $2458558.60961$  & $ 0.0\pm 1.0$ & TESS \\
2734 & $2458559.39918$  & $ 1.0\pm 1.0$ & TESS \\
2735 & $2458560.18695$  & $ -0.5\pm 1.0$ & TESS \\
2736 & $2458560.97701$  & $ 1.3\pm 1.0$ & TESS \\
2737 & $2458561.76565$  & $ 1.0\pm 1.0$ & TESS \\
2738 & $2458562.55360$  & $ -0.3\pm 1.0$ & TESS \\
2739 & $2458563.34290$  & $ 0.4\pm 1.0$ & TESS \\
2740 & $2458564.13196$  & $ 0.7\pm 0.9$ & TESS \\
2741 & $2458564.92117$  & $ 1.2\pm 1.0$ & TESS \\
2742 & $2458565.70927$  & $ 0.1\pm 1.0$ & TESS \\
2743 & $2458566.49876$  & $ 1.1\pm 0.9$ & TESS \\
2744 & $2458567.28769$  & $ 1.2\pm 1.0$ & TESS \\
2745 & $2458568.07643$  & $ 1.0\pm 1.0$ & TESS \\

\hline
\end{longtable} 
}

\longtab{
\begin{longtable}{cccc}
\caption{\label{times_wasp77}Transit mid-times of WASP-77Ab}\\
\hline \hline
Epoch & Transit mid-time & TTV & References\\
      & (${\rm BJD_{TDB}}$) & (min) &  \\
\hline
\endfirsthead
\caption{Transit mid-times of WASP-77Ab continued}\\
\hline\hline
Epoch & Transit mid-time & TTV & References\\
      & (${\rm BJD_{TDB}}$) & (min) &  \\
\hline
\endhead
\hline
\endfoot
-1140 & $2455870.45054$ & $-2.0\pm1.5$ & \citet{Maxted2013} \\
-845 & $2456271.65888$ & $-2.0\pm1.4$ & \cite{Turner2016} \\
-659 & $2456524.62617$ & $0.8\pm1.8$ & Exoplanet Transit Database \\
-606 & $2456596.70591$ & $-1.7\pm1.5$  & Exoplanet Transit Database  \\
-92 & $2457295.7626$  & $1.2\pm1.3$& TraMoS \\
-89 & $2457299.84119$ &  $-1.0\pm1.3$ & TraMoS \\
175 & $2457658.88744$ & $-2.8\pm1.5$ & TraMoS\\
177 & $2457661.60987$   & $0.6\pm1.7$ & Exoplanet Transit Database \\
183 & $2457669.77054$ &  $1.4\pm1.4$  & TraMoS  \\
229 & $2457732.33382$ & $4.2\pm1.4$& Exoplanet Transit Database \\
447 & $2458028.8159$  & $-1.8\pm6.5$ & TraMoS\\ 
728 & $ 2458410.98440$  & $ -1.2\pm 1.3$ & TESS \\
729 & $2458412.34460$  & $ -1.0\pm 1.3$ & TESS \\
730 & $2458413.70531$  & $ 0.0\pm 1.3$ & TESS \\
731 & $2458415.06491$  & $ -0.6\pm 1.3$ & TESS \\
732 & $2458416.42497$  & $ -0.6\pm 1.3$ & TESS \\
733 & $2458417.78493$  & $ -0.7\pm 1.3$ & TESS \\
736 & $2458421.86480$  & $ -1.0\pm 1.3$ & TESS \\
739 & $2458425.94511$  & $ -0.7\pm 1.3$ & TESS \\
740 & $2458427.30511$  & $ -0.7\pm 1.3$ & TESS \\
741 & $2458428.66460$  & $ -1.5\pm 1.3$ & TESS \\
742 & $2458430.02488$  & $ -1.1\pm 1.3$ & TESS \\
743 & $2458431.38494$  & $ -1.1\pm 1.3$ & TESS \\
744 & $2458432.74541$  & $ -0.4\pm 1.3$ & TESS \\
745 & $2458434.10526$  & $ -0.7\pm 1.3$ & TESS \\
746 & $2458435.46538$  & $ -0.6\pm 1.3$ & TESS \\
\hline
\end{longtable}
}

\twocolumn

\end{document}